\documentclass[nonblindrev]{informs3noheader}

\theoremstyle{EX}

\usepackage{enumitem}
\usepackage{thm-restate}

\usepackage{tikz}
\usetikzlibrary{calc}
\usetikzlibrary{decorations.pathreplacing}
\usetikzlibrary{decorations.markings}
\usetikzlibrary{arrows}

\newcommand{\R}{\mathbb{R}}

\newcommand{\mb}[1]{\ensuremath{\boldsymbol{#1}}}

\newcommand{\cK}{\mathcal{K}}

\newcommand{\Ocal}{\mathcal{O}}

\renewcommand{\epsilon}{\varepsilon}

\newcommand{\comment}[1]{}

\usepackage{natbib}
 \NatBibNumeric
 \def\bibfont{\small}%
 \bibpunct[, ]{[}{]}{,}{n}{}{,}%

\usepackage[colorlinks=true,breaklinks=true,bookmarks=true,urlcolor=blue,
     citecolor=blue,linkcolor=blue,bookmarksopen=false,draft=false]{hyperref}

\definecolor{mygreen}    {RGB}{0,90,0}
\definecolor{myblue}     {RGB}{0,51,140}
\definecolor{myorange}   {RGB}{238,118,0}
\definecolor{myred}      {RGB}{126,0,0}
\definecolor{mygray}     {RGB}{100,100,105}
\definecolor{mygrayblue} {RGB}{0,128,128}
\definecolor{mygraygreen}{RGB}{128,128,0}
\definecolor{DarkPurple}     {RGB}{142, 36, 170}
\definecolor{LightPurple}    {RGB}{57, 130, 7}

\def\EMAIL#1{\href{mailto:#1}{#1}}

\TheoremsNumberedThrough
\EquationsNumberedThrough

\begin{document}
 \RUNAUTHOR{Blanchard, Jacquillat, and Jaillet}
 \RUNTITLE{Probabilistic bounds on the TRP and $k$-TSP}
\TITLE{Probabilistic bounds on the $k-$Traveling Salesman Problem and the Traveling Repairman Problem}
\ARTICLEAUTHORS{
\AUTHOR{Mo\"ise Blanchard, Alexandre Jacquillat, Patrick Jaillet}
\AFF{Massachusetts Institute of Technology, \EMAIL{moiseb@mit.edu,alexjacq@mit.edu,jaillet@mit.edu}}
}

\ABSTRACT{
The $k-$traveling salesman problem ($k$-TSP) seeks a tour of minimal length that visits a subset of $k\leq n$ points. The traveling repairman problem (TRP) seeks a complete tour with minimal latency. This paper provides constant-factor probabilistic approximations of both problems. We first show that the optimal length of the $k$-TSP path grows at a rate of $\Theta\left(k/n^{\frac{1}{2}\left(1+\frac{1}{k-1}\right)}\right)$. The proof provides a constant-factor approximation scheme, which solves a TSP in a high-concentration zone---leveraging large deviations of local concentrations. Then, we show that the optimal TRP latency grows at a rate of $\Theta(n\sqrt n)$. This result extends the classical Beardwood-Halton-Hammersley theorem to the TRP. Again, the proof provides a constant-factor approximation scheme, which visits zones by decreasing order of probability density. We discuss practical implications of this result in the design of transportation and logistics systems. Finally, we propose dedicated notions of fairness---randomized population-based fairness for the $k$-TSP and geographical fairness for the TRP---and give algorithms to balance efficiency and fairness.
}


\maketitle

\section{Introduction}

This paper studies the traveling repairman problem (TRP)---also known as the minimum latency problem \citep{afrati1986complexity, minieka1989delivery,bianco1993traveling}---and the $k-$traveling salesman problem ($k$-TSP) in the Euclidean plane. These two problems are extensions of the well-studied traveling salesman problem (TSP). The TSP takes as inputs a set of $n$ points as well as a distance matrix between all points, and seeks the route of minimal length that visits all $n$ points. Assuming constant speed, the TSP is equivalent to minimizing the arrival time at the end of the tour. Instead, the TRP seeks a tour that minimizes the sum of waiting times, known as the total latency. This problem arises in routing problems with requirements on customer wait times, for instance, to ensure sufficient level of service, or to maximize operating profitability under random customer abandonment. The TRP is also applicable to disk head scheduling \citep{blum1994minimum}, flexible manufacturing systems \citep{simchi1991minimizing}, machine scheduling \citep{picard1978time}, information search in computer networks \citep{ausiello2000salesmen} and others domains \citep{tsitsiklis1992special}.

In contrast, the $k$-TSP seeks a path of minimal length that visits $k$ out of $n$ points, where $k\leq n$. In other words, the server chooses which points to serve. This problem has natural applications in routing and distribution systems, e.g., for a logistics provider that can only serve a partial set of customers due to limitations on its delivery capacity. In addition, the $k$-TSP has been used as subroutine for TRP approximation algorithms \citep{blum1994minimum,goemans1998improved}.

Our goal is to derive probabilistic bounds on the optimal $k$-TSP tour and the optimal TRP latency, which, in turn, lead to the design of efficient probabilistic approximation schemes. We consider a setting with a fixed number $n$ of points in the Euclidean plane. The location of these points is unknown, following a known distribution---we denote by $f$ the density of its absolutely-continuous part. We seek \emph{constant-factor optimal} approximations, that is, probabilistic solutions leading to an objective value that is asymptotically within a constant factor from the optimal solution. Specifically, we derive constant-factor estimates for the $k$-TSP and TRP solutions as a function of the number of points $n$ and the density $f$. Moreover, through constructive proofs, we provide constant-factor approximation algorithms for both problems.

\subsection{Related work}

The Traveling Salesman Problem (TSP) is one of the canonical problems in operations research. The Beardwood-Halton-Hammersley theorem, stated in \citep{beardwood1959shortest} and improved in \citep{steele1981subadditive,steele1997probability} gives a constant-factor $\Theta(\sqrt{n})$ approximation of the optimal TSP tour in the Euclidean space. The proof of these TSP estimates leads to the design of approximation algorithms that are stochastically robust in the a priori setting. A priori optimization \citep{bertsimas1990priori} provides an optimization framework when the same combinatorial problem is solved repeatedly over different instances. The goal is to compute a master solution ahead of time that minimizes an expected cost function, given subsequent adjustments according to simple rules upon the realization of uncertainty.

This work has leveraged extensively the ``locality property'' of the TSP to design ``divide and conquer" approximation algorithms. That is, under this approach, we define an a priori route that can then be slightly modified to respond to the instance realizations, while keeping its approximation guarantees \citep{carlsson2018coordinated}. Moreover, a near-optimal tour for the TSP objective remains near-optimal if we change the starting point of the tour. Even in the case of a unique starting depot, restricting the server to start serving from any point in the tour only induces an additional constant cost (which does not scale up with the number of points).

Despite its similarity with the TSP, the TRP lacks a locality property, and is therefore much harder to solve. Local changes in the input points affect the waiting time of all the remaining ones, leading to non-local modifications in the optimal tour. Even in the one-dimensional case where points lie on a line, the optimal TRP tour may cross itself several times, which is not the case in the TSP. \citet{blum1994minimum} showed that there exists a simple reduction from the TRP to the TSP, implying that the TRP is NP-hard in general for all metric spaces where the TSP is known to be NP-hard. The TRP is even NP-hard on weighted trees, where the TSP is easy \citep{sitters2002minimum}.

\citet{blum1994minimum} proposed the first constant-factor approximation algorithm for the TRP in general metric spaces. Their approach involves a reduction to the $k-$Minimum Spanning Tree ($k-$MST) problem, which seeks an optimal tree spanning $k$ vertices in a weighted graph. This problem is also known to be NP-hard \citep{fischetti1994weighted}. Substantial work has been made to give approximation algorithms for this problem \citep{ravi1996spanning,blum1996constant,garg19963,arya19982}, with the current best bound being a $2-$approximation algorithm \cite{garg2005saving}. More precisely, \citet{blum1994minimum} showed that a $c-$approximating algorithm for $k-$MST yields an $8c-$approximating algorithm for the TRP, thus providing a $16-$approximation using the best-known algorithm for the $k-$MST. \citet{goemans1998improved} improved the reduction in \citep{blum1994minimum} from a factor of $8$ to a factor of $3.59$. \citet{chaudhuri2003paths} gave the current best bound, a $3.59-$approximation algorithm for the TRP in general metric spaces. In the case of weighted trees on the Euclidean plane, there exists a polynomial time $(1+\epsilon)-$approximation algorithm \citep{sitters2014polynomial}.

The $k-$MST and $k$-TSP are also closely related. Hence, some papers on the $k-$MST give results for the $k$-TSP. Specifically, the algorithms given by \citet{blum1996constant}, \citet{garg19963,garg2005saving} and \citet{arora20062+} can be adapted to the $k$-TSP, which yields a $2-$approximation algorithm for the $k$-TSP. These results have also been leveraged to address other variants, such as prize-collector problems \citep{johnson2000prize,paul2020budgeted}. More recently, \citet{pandiri2020two} gave metaheuristics for the rooted $k$-TSP leveraging permutation-based and local-search heuristics.

Recent work has focused on the a priori TRP \citep{van2018priori,navidi2020approximation}. Following earlier work on the a priori TSP \citep{bertsimas1990priori,jaillet1993analysis,laporte1994priori}, this problem seeks a master tour under demand uncertainty, where each vertex is present with some probability. In this paper, we seek a priori solutions when the uncertainty lies in the position of the points, as opposed to the number of such points.

Unlike the TSP, the $k$-TSP and the TRP encode a notion of priority between points. In the $k$-TSP, the decision-maker can choose which points to serve; in the TRP, the decision-maker can choose the sequence of customer visits. Such prioritization gives rise to important fairness issues. Namely, in the $k$-TSP, one can serve the points that lie in high-density zones, ignoring all other points altogether. Similarly, in the TRP, one can serve zones by decreasing order of density, thus prioritizing points in high-density zones over points in low-density zones. As a result, the approximation algorithms for both problems can lead to spatial discrimination across populations. This trade-off between efficiency and fairness arises in many resource allocation and scheduling problems \citep{bertsimas2011price,bertsimas2012efficiency}, spanning communication networks \citep{luo2004packet,radunovic2007unified,bertsimas2011price}, air traffic management \citep{vossen2003general,bertsimas2011proposal,jacquillat2018interairline} and finance \citep{o2006pooling}.

\subsection{Contributions and outline.}

This paper makes three contributions:
\begin{itemize}[leftmargin=*]
    \item[--] We derive a constant-factor probabilistic estimate of the optimal $k$-TSP tour for general distributions (Section~\ref{section:kTSP}). Specifically, we show that the optimal $k$-TSP length grows at a rate of $\Theta\left(k/n^{\frac{1}{2}\left(1+\frac{1}{k-1}\right)}\right)$. This result is obtained by leveraging large deviations in local point concentration to serve regions with high point concentration (especially for small $k$).
    
    \item[--] We provide non-asymptotic constant-factor estimates of the optimal TRP for general distributions (Section~\ref{section:TRP}). We show that total latency grows as $\Theta(n\sqrt n)$ and characterize the dependence of the constant on the sampling distribution as the integral of a function of absolutely-continuous-part density---thus extending the BHH result from the TSP to the TRP. We discuss practical implications for the design of transportation and logistics systems in Section~\ref{subsec:logistics}.
    
    \item[--] We define fairness-enhanced versions of the $k$-TSP and TRP, and analyze the price of fairness (Section~\ref{section:fairness}). The approximation algorithms for the $k$-TSP and the TRP are highly ``local''. As a result, customers in high-density regions are more likely to receive a service (for the $k$-TSP) or to have a lower wait time (for the TRP). We define notions of fairness to circumvent this issue. For the TRP, we show that our approximation scheme satisfies max-min fairness, and propose modifications toward proportional fairness. For the $k$-TSP, we show that geographical fairness across regions leads to significant efficiency loss. We thus propose population-based fairness, given the distribution of populations across regions. We show that probabilistic population-based fairness still allows for flexibility, and can lead to near-optimal $k$-TSP solutions.
    
\end{itemize}

Before proceeding, we first describe in Section~\ref{sec:preliminaries} the modeling framework, outline our main results along with the proof techniques, and discuss their practical implications.

\section{Setup, overview of results, and practical implications}\label{sec:preliminaries}

\subsection{Setup and preliminaries}

We consider a set of $n$ points $V=\{X_1,\ldots,X_n\}$ in the Euclidean space $\R^2$ equipped with the natural Euclidean distance. We focus on the 2-dimensional case, but our results can easily be extended to the general case $\R^d$. We consider a probabilistic setting where vertices $X_1,\ldots,X_n$ are independent and identically distributed, drawn from some distribution on a compact $\cK\subset \R^2$. We denote by $f$ the density of its absolutely-continuous part.
 
Given the set of points $V$, we consider three optimization problems:
\begin{enumerate}
\item The traveling salesman problem (TSP) seeks a tour that starts in a vertex, visits all $n$ vertices with some service order $x_1,\ldots,x_n$, and returns to the starting point. The objective is to minimize the total length of the tour:
\begin{equation}\label{eq:total_length}
    \sum_{i=1}^{n-1}|x_{i+1}-x_i|+|x_1-x_n|.
\end{equation}
\item The $k$-traveling salesman problem ($k$-TSP), which seeks a path that visits an endogenous subset of $k\leq n$ vertices $x_1,\ldots,x_k$. The objective is again to minimize the total length of the path:
\begin{equation}\label{eq:k_length}
    \sum_{i=1}^{k-1}|x_{i+1}-x_i|.
\end{equation}
\item The traveling repairman problem (TRP). Like the TSP, the TRP also seeks a complete tour of the $n$ vertices. However, the TRP minimizes the total latency, or the total wait times at the vertices. Formally, if $x_1,\cdots,x_n$ defines a service order, the latency at point $x_i$ is defined as $l_i = \sum_{j=1}^{i-1} |x_{j+1}-x_j|.$ The TRP tour minimizes the sum of latencies:
\begin{equation}\label{eq:total_latency}
    \sum_{i=1}^n l_i = \sum_{i=1}^{n-1} (n-i)|x_{i+1}-x_i|.
\end{equation}
\end{enumerate}

In this paper, we provide \emph{constant-factor} probabilistic bounds, i.e., bounds on the expected optimal value of these problems that hold asymptotically within a universal constant factor, where the expectation is taken over the randomness of the points $X_1,\ldots, X_n$ (our bounds also hold with high probability). Similarly, we say that an algorithm is \emph{constant-factor optimal} if it provides solutions with objective value within a constant factor of the optimal solution in expectation.

In this setting, the well-known BHH theorem shows that the optimal TSP length grows as $\Theta(\sqrt{n})$.

\begin{theorem}[BHH theorem, \citet{beardwood1959shortest}]
\label{thm:BHH}
Let $(X_i)_{i\geq 1}$ be a sequence of i.i.d. random points according to a distribution on a compact space $\cK\subset \R^2$. With probability one, the length $l_{TSP}(X_1,\ldots,X_n)$ of the optimal TSP on points $\{X_1.\ldots,X_n\}$ satisfies
\begin{equation*}
    \lim_{n\to\infty} \frac{l_{TSP}(X_1,\ldots,X_n)}{\sqrt n} =\beta_{TSP} \iint_\cK \sqrt{f(x)}dx,
\end{equation*}
where $0.6250\leq \beta_{TSP}\leq 0.9204$ is a universal constant and $f$ denotes the density of the absolutely-continuous part of the distribution.
\end{theorem}

Lemma~\ref{lemma:upper bound tsp} provides a simplified version of Theorem~\ref{thm:BHH} that will be useful in our analysis. The proof of this result constructs a simple ``master" space-filling curve that is at most $\frac{1}{2\sqrt n}$ away from any point in the unit square and has length $\sqrt n+\Ocal(1)$. We can adapt this simple curve to serve any vertex by adding a ``back-and-forth" detour from the closest point on the curve. Similarly, we can adapt the curve to serve $n$ points. The length of the resulting tour is $2\sqrt n + \Ocal(1)$.

\begin{lemma}[\citet{beardwood1959shortest}]
\label{lemma:upper bound tsp}
Let $n\geq 2$ and $(X_i)_{1\leq i\leq n}$ points in the unit square $[0,1]^2$. Denote by $l_{TSP}(X_1,\ldots,X_n)$ the length of the TSP tour visiting these points. Then,
\begin{equation*}
    l_{TSP}(X_1,\ldots,X_n)\leq 2\sqrt{n} + C,
\end{equation*}
for some universal constant $C>0$.
\end{lemma}

In our algorithms for the $k$-TSP and the TRP, we will use this result as a subroutine, to design an a priori curve that can serve $n$ points with a worst-case length of $2\sqrt n +C$. Asymptotically, this a priori procedure yields solutions that are at most $2/\beta_{TSP}$-away from the optimal TSP tour.

\subsection{Main results}

The main results of the paper provide constant-factor approximations of the $k$-TSP and TRP solutions. First, we show in Section~\ref{section:kTSP} that the optimal $k$-TSP tour grows at a rate of $\Theta\left(k/n^{\frac{1}{2}\left(1+\frac{1}{k-1}\right)}\right)$ (Theorem~\ref{thm:exact rate kTSP}). This rate can be interpreted as a positive result by contrasting it with (i) a naive bound of $\sqrt{k}$, which applies a TSP tour on a random subset of $k$ points; and (ii) a bound of $\frac{k}{\sqrt{n}}$, which selects the best subpath of $k$ consecutive vertices in the full TSP tour. The rate of $\Theta\left(k/n^{\frac{1}{2}\left(1+\frac{1}{k-1}\right)}\right)$ underscores a benefit of $\sqrt{k/n}$ that comes from merely optimizing which vertices to visit and an additional benefit of $n^{{-\frac{1}{2(k-1)}}}$ that comes from re-optimizing the tour---leveraging large deviations in local point concentration.

The proof of the $k$-TSP proceeds by showing that the rate $k/n^{\frac{1}{2}\left(1+\frac{1}{k-1}\right)}$ is non-asymptotically tight up to a constant with uniform densities. We extend the analysis to the case of general measurable (not necessarily continuous) densities. In particular, the proof for the upper bound is constructive, and provides a constant-factor approximation algorithm when $1\ll k\ll n$, by selecting the region with highest point concentration and performing the (uniform) $k$-TSP in this region.

Second, we show in Section~\ref{section:TRP} that the optimal TRP latency grows at a rate of $\Theta\left(n\sqrt{n}\right)$ (Theorem~\ref{thm:asymptotic bound TRP}). In contrast to the previous one, this is a rather negative result. Indeed, the TSP tour gives a $\Theta(\sqrt{n})$ estimate of the latency in the last vertex. Accordingly, if all customers had to wait as long as the \emph{last} customer, we would end up with a total latency of the order of $n\sqrt{n}$. As this result shows, even by re-optimizing the tour, the TRP still leads to optimal latency on the order of $n\sqrt{n}$.

The proof of the TRP upper bound is also constructive and gives a simple constant-factor approximation scheme. This scheme constructs a ``master a priori tour" depending solely on the absolutely-continuous-part density, then adapts it to any realization of sampled points. Specifically, the algorithm partitions the region into zones of constant density, visits zones by decreasing order of local density, and performs a tour on each zone following space-filling techniques for the TSP.

From a practical standpoint, the TRP result is structurally different from the TSP result. Specifically, the optimal TSP tour is concave in the number of vertices, indicating economies of scale. In contrast, the optimal TRP latency is convex in the number of vertices, indicating diseconomies of scale. This distinction has implications for the design of transportation and logistics systems.

\subsection{Implications for transportation and logistics operations}\label{subsec:logistics}

TSP approximation results provide insights into the operations of transportation and logistics systems, which can be used to support upstream planning decisions. Sample applications include location analysis \citep{carlsson2022continuous}, area partitioning for vehicle routing \cite{carlsson2012dividing} and same-day delivery systems \cite{stroh2022tactical,banerjee2022has,banerjee2022fleet}. In these problems, continuous approximations estimate routing costs into upstream optimization models---rather than, for instance, capturing discrete routing dynamics at significant computational costs.

Specifically, TSP approximation results take the perspective of a logistics provider. However, several systems strive to also minimize customer wait times. For instance, in food delivery, a company needs to serve customers as early as possible as opposed to meeting an overall deadline. As another example, school bus (or company bus) routing aims to minimize the travel times of the students (or employees), as opposed to the vehicle's trip time. The TRP provides the natural framework to estimate customer level-of-service. As such, the results of this paper can be used to guide the design of such transportation and logistics systems focused on wait times.

This distinction between the TSP length and the TRP latency has practical consequences due to the concavity of the $\sqrt{n}$ function versus the convexity of the $n\sqrt{n}$ function. As a result, economies of scale in the TSP favor service \emph{concentration} (few vehicles, each serving many customers), whereas diseconomies of scale in the TRP favor service \emph{dispersion} (more vehicles each serving a smaller number of customers). We illustrate this tension below in two simple examples.

\paragraph{Fleet size optimization.}
We seek the number of vehicles $m$ to serve a batch of $N$ orders. Each vehicle incurs a fixed cost $c$ and each vehicle carries $\frac{N}{m}$ orders. Assume first that the system minimizes vehicles' fixed costs and travel costs. Based on the BHH approximation, we can write this objective as minimizing $ c\cdot m + d\cdot m\cdot \sqrt{\frac{N}{m}} = cm+d\sqrt{Nm}$, for some scaling constant $d$. The optimal strategy is $m=1$, even with $c=0$, that is, a single vehicle serves all customers. However, if we replace the vehicle travel time component with a customer wait time component, the objective becomes minimizing $c\cdot m + \tilde d\cdot \frac{N}{m}\sqrt{\frac{N}{m}}$, for some scaling constant $\tilde d$. The optimum is now attained for $m^* = \left(\frac{3\tilde d}{2c}\right)^{2/5} N^{3/5}$. Now, the operator leverages a multi-vehicle fleet, which increases with customer demand. This example underscores two opposite strategies, spanning pure consolidation in the TSP case (serving the entire batch with a single vehicle) versus dispersion in the TRP case (serving customer demand with multiple vehicles to balance vehicle costs and customer wait times).

\begin{figure}[h!]
        \centering
        \begin{tikzpicture}[scale=0.7]
        \draw[color=black,thick] (-0.5,0) to (20,0);
        \draw[color=black,thick] (-0.5,-0.2) node[below]{$0$} -- (-0.5,0.2);
        \draw[color=black,thick] (20,-0.2) node[below]{$T$} -- (20,0.2);
        \draw[color=black,thick] (15,-0.2) node[below]{$T_{cutoff}$} -- (15,0.2);
        \foreach \i in {0,...,15}{
            \draw[color=myred,fill=myred] (\i*.5,0) circle [radius=0.1];
        }
        \foreach \i in {16,...,25}{
            \draw[color=myblue,fill=myblue] (\i*.5,0) circle [radius=0.1];
        }
        \foreach \i in {26,...,29}{
            \draw[color=mygreen,fill=mygreen] (\i*.5,0) circle [radius=0.1];
        }
        \draw[-latex,color=myred, ultra thick, bend left = 25 pt] (7.5,0) to (20,0);
        \draw[-latex,color=myblue, ultra thick, bend right = 30 pt] (12.5,0) to (20,0);
        \draw[-latex,color=mygreen, ultra thick, bend left = 20 pt] (14.5,0) to (18,0);
        \end{tikzpicture}
        \begin{tikzpicture}[scale=0.7]
        \draw[color=black,thick] (-0.5,0) to (20,0);
        \draw[color=black,thick] (-0.5,-0.2) node[below]{$0$} -- (-0.5,0.2);
        \draw[color=black,thick] (20,-0.2) node[below]{$T$} -- (20,0.2);
        \draw[color=black,thick] (15,-0.2) node[below]{$T_{cutoff}$} -- (15,0.2);
        \foreach \i in {0,...,7}{
            \draw[color=myred,fill=myred] (\i*.5,0) circle [radius=0.1];
        }
        \foreach \i in {8,...,14}{
            \draw[color=myblue,fill=myblue] (\i*.5,0) circle [radius=0.1];
        }
        \foreach \i in {15,...,22}{
            \draw[color=mygreen,fill=mygreen] (\i*.5,0) circle [radius=0.1];
        }
        \foreach \i in {23,...,29}{
            \draw[color=DarkPurple,fill=DarkPurple] (\i*.5,0) circle [radius=0.1];
        }
        \draw[-latex,color=myred, ultra thick, bend left = 30 pt] (3.5,0) to (8.5,0);
        \draw[-latex,color=myblue, ultra thick, bend right = 30 pt] (7,0) to (12,0);
        \draw[-latex,color=mygreen, ultra thick, bend left = 20 pt] (11,0) to (16,0);
        \draw[-latex,color=DarkPurple, ultra thick, bend right = 20 pt] (14.5,0) to (19.5,0);
        \end{tikzpicture}
        \caption{Consolidation-driven dispatch based on order deadlines from the TSP approximation (top), versus dispersion-driven dispatch based on customer wait times from the TRP approximation (bottom) for $m=4$ vehicles.}
        \label{fig:TSPTRP}
    \end{figure}
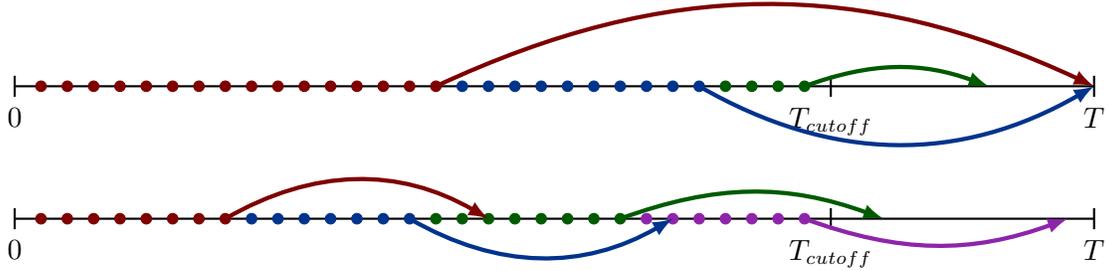

\paragraph{Vehicle dispatch in same-day-delivery (SDD) systems.}
Based on \cite{stroh2022tactical}, we consider an SDD provider that operates a fleet of $m$ vehicles, each of which can only be dispatched once. Customers arrive at a constant rate $\lambda$ until an order cutoff $N$ is met at time $T_{cutoff}=N/\lambda$. The operator optimizes dispatch decisions, characterized by a dispatch time $t_i$ and a number of carried orders $n_i$ for each vehicle $i=1,\cdots,m$. Following the BHH approximation, the delivery time of vehicle $i$ can be written as $a\cdot\sqrt n_i$ for some scaling constant $a$. The SDD constraint asks that vehicles should complete their deliveries by an end-of-day deadline $T$, that is, $t_{i} + a\sqrt{n_{i}}\leq T$ for all $i=1,\cdots,m$. \citet{stroh2022tactical} minimize the total dispatch time $\sum_{i=1}^m a \sqrt{n_{i}}$ under the aforementioned SDD constraints, demand constraints (all orders need to be served), and consistency constraints (orders can only be carried after they become available). Whenever feasible, the optimal strategy is to dispatch the first vehicle when it can fulfill all revealed orders and return exactly at time $T$; the second vehicle when it can fulfill all subsequent orders and return exactly at time $T$; etc. (top of Figure~\ref{fig:TSPTRP}). This strategy is feasible (hence, optimal), whenever the fleet $m$ is sufficiently large to cover all the demand, which can be checked by solving recursively the equations $t_i+a\sqrt{\lambda(t_i-t_{i-1})}=T$ for $t_{i-1}\leq t_i\leq T$ with $t_0=0$ and checking whether $t_m\geq T_{cutoff}$.

Now assume that the operator minimizes customer wait times. Based on our TRP approximation result, this scales as $w\cdot n\sqrt{n}$ for some scaling constant $w$. Note that the cost function can be augmented by replacing $wn\sqrt{n}$ with $b\cdot n^2+w\cdot n\sqrt{n}$, where $b\cdot n^2$ captures the batching time prior to the dispatch and $w\cdot n\sqrt{n}$ captures the wait time after the dispatch. Either way, the cost function is now convex in $n$. Whenever feasible, the optimal strategy is therefore to dispatch vehicles at regular times $\frac{iN}{m\lambda}$ (bottom of Figure~\ref{fig:TSPTRP}). This strategy is feasible (hence, optimal), whenever the last vehicle $m$ can complete its orders by the end of the day, i.e., whenever $\frac{N}{\lambda} + a\sqrt{\frac{N}{m}} \leq T$.

Again, this structure underscores two opposite strategies. In the SDD system (based on a TSP objective), the dispatching policy leverages consolidation, by bundling orders together as much as possible. In contrast, in the food delivery, school bus and employee bus systems, the dispatching policy leverages dispersion, by distributing orders as evenly as possible across possible. Although stylized, these two examples underscore that minimizing wait times may significantly alter design decisions in routing systems, as compared to focusing on vehicle travel times.

\section{The \texorpdfstring{$k$}{k}-Traveling Salesman Problem}
\label{section:kTSP}

We provide probabilistic estimates on the length of the $k$-TSP tour. Before proceeding, let us expand on the two aforementioned naive bounds:
\begin{itemize}
    \item[--] Upper bound of $\Ocal(\sqrt k)$: By choosing the $k$ points to visit uniformly at random among the $n$ available points, the BHH theorem ensures that the length of the optimal path visiting these $k$ points has length $\sim \beta_{TSP} \sqrt k \iint_\cK \sqrt f$ as $k\to\infty$. However, this analysis does not leverage the flexibility regarding which points to serve.
    \item[--] Upper bound of $\Ocal\left(\frac{k}{\sqrt{n}}\right)$: Consider the optimal TSP tour visiting all $n$ points of length $l_{TSP}(n)$. Selecting $k$ consecutive points on this tour at random---we randomly select the starting point---yields a path of length $\frac{k-1}{n}l_{TSP}(n)$ in expectation. In particular, the best choice of $k$ consecutive points on the TSP tour yields an upper bound for the $k$-TSP of $\frac{k-1}{n}l_{TSP}(n) = \Ocal(k/\sqrt n)$. This observation underscores the benefits of choosing which points to serve. As we shall see, such flexibility can be very significant, especially for small values of $k$. Yet, this analysis still relies on the optimal TSP tour, therefore eliminating an extra degree of freedom in the $k$-TSP.
\end{itemize}
We will show that this rate $\Ocal(k/\sqrt n)$ is essentially tight for large $k$, but can be tightened for small $k$. For instance, for $k=2$, the minimum distance between $n$ uniformly sampled points in the unit square is $\Theta(1/n)$ instead of $\Ocal(1/\sqrt n)$. Our results in this section interpolate the $\Theta(1/n)$ estimate for $k=2$ and the $\Theta(\sqrt n)$ estimate for $k=n$. We now present the main result of this section giving the exact rate of the expected $k$-TSP length. Note that this result does not only provide an asymptotic rate, but holds yields an estimate of the $k$-TSP length for any choice of $2\leq k\leq n$.

\begin{theorem}
\label{thm:exact rate kTSP}
Assume $n$ vertices are drawn independently, uniformly on a compact space $\cK\subset \mathbb R^2$ with area $\mathcal A_\cK$. Denote by $l_{TSP}(k,n)$ the length of the $k$-TSP on these $n$ vertices. Then, for all $n\geq 2$ and $2\leq k\leq n$, for some universal constants $0<c<C$,
\begin{equation*}
    c\frac{k-1}{n^{\frac{1}{2}\left(1+\frac{1}{k-1}\right)}}\sqrt{\mathcal A_\cK} \leq \mathbb E[l_{TSP}(k,n)]\leq C\frac{k-1}{n^{\frac{1}{2}\left(1+\frac{1}{k-1}\right)}}\sqrt{\mathcal A_\cK}.
\end{equation*}
\end{theorem}

Theorem~\ref{thm:exact rate kTSP} exhibits an additional factor $\Theta(n^{-\frac{1}{2(k-1)}})$ compared to the previous bound $\Ocal(k/\sqrt n)$. This additional factor corresponds to large deviations of local point densities. Consider any sub-square of area $\Ocal(k/n)$, and perform the TSP on this sub-square. We would expect $\Ocal(k)$ points in this subsquare, yielding a path of length $\Ocal(\sqrt k \cdot \sqrt{k/n})=\Ocal(k/\sqrt n)$. In the $k$-TSP however, we can choose to serve zones with abnormally-high point concentration---deviating from the expected density. In the following two subsections we prove Theorem~\ref{thm:exact rate kTSP} and show that the resulting discount on the length of the optimal path visiting $k$ points is the additional factor $\Theta(n^{-\frac{1}{2(k-1)}})$.

\subsection{Lower bounds on the \texorpdfstring{$k$}{k}-TSP}
\label{subsection:lower bound kTSP}
We will first need the following lemma.

\begin{restatable}{lemma}{LemmakTSPEstimate}
\label{lemma:k-tsp estimate}
Assume all $n$ vertices are drawn independently, uniformly on a compact $\cK\subset \mathbb R^2$ with area $\mathcal A_\cK$. Denote by $l_{TSP}(k,n)$ the length of the $k$-TSP on these $n$ vertices. Then, for any $\alpha>0$,
\begin{equation*}
    \mathbb P\left[l_{TSP}(k,n)\leq \alpha\right] \leq n^k \left(\frac{2\pi\alpha^2}{\mathcal A_\cK}\right)^{k-1}\frac{1}{(2k-2)!}.
\end{equation*}
\end{restatable}

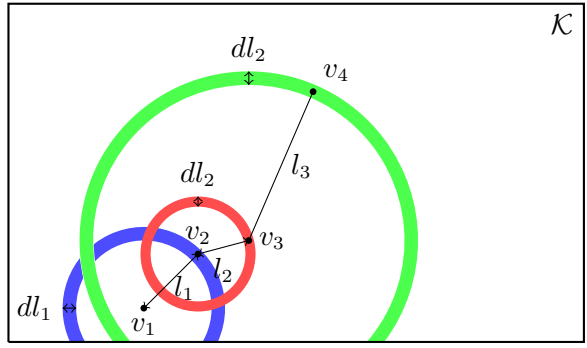
\begin{figure}
    \centering
    \begin{tikzpicture}[scale=0.9]
    \coordinate (A) at (-2.5,-1.5);
    
    \fill (A) circle(0.05);
    \draw (A) node[below] {$v_1$};
    
    \draw[white] (A) ++ (-30:1) node (A1) {} arc(-30:180+30:1) node (A2) {};
    \draw[white] (A) ++ (-25:1.2) node (A3) {} arc(-25:180+25:1.2) node (A4) {};
    
    \fill [blue!70] (A1) arc(-30:180+30:1) -- (A4) arc(180+25:-25:1.2) -- (A1);
    
    \coordinate (B) at (-1.7,-0.7);
    \coordinate (B1) at (-1,-0.7);
    \coordinate (B2) at (-0.85,-0.7);
    
    \fill (B) circle(0.05);
    \draw (B) node[above] {$v_2$};
    
    \fill[red!70,even odd rule] (B) circle(0.7) circle(0.85);
   
    \coordinate (C) at (-0.95,-0.5);
    \fill (C) circle(0.05);
    \draw (C) node[right] {$v_3$};
    
    \draw (C)[white] ++ (-40.5:2.3) node (C1) {} arc(-40.5:180+40.5:2.3) node (C2) {};
    \draw (C)[white] ++ (-37:2.5) node (C3) {} arc(-37:180+37:2.5) node (C4) {};
    
    \fill [green!70] (C1) arc(-40.5:180+40.5:2.3) -- (C4) arc(180+37:-37:2.5) -- (C1);
    
    \coordinate (D) at (0,1.7);
    \fill (D) circle(0.05);
    \draw (D) node[above right] {$v_4$};
        
    \draw[<->] (A)-- (B);
    \draw (-1.9,-1.2) node {$l_1$};
    \draw[<->] (B)-- node[below] {$l_2$} (C);
    \draw (C) -- node[right] {$l_3$} (D);
    
    \draw[<->] (A) ++(180:1) -- ++(180:0.2)  node[left] {$dl_1$};
    \draw[<->] (B) ++(90:0.7) -- ++(90:0.15)  node[above] {$dl_2$};
    \draw[<->] (C) ++(90:2.3) -- ++(90:0.2)  node[above] {$dl_2$};
    
    \draw[thick] (-4.5,-2) -- (4,-2) -- (4,3) node[below left] {$\cK$} --(-4.5,3) -- (-4.5,-2);
    \end{tikzpicture}
    \caption{Illustration of the proof of Lemma~\ref{lemma:k-tsp estimate}: $\mathbb{P}\left(l_i\leq |v_{i+1}-v_i|\leq l_i+dl_i\right)\leq\frac{2\pi l_i}{\mathcal A_\cK}dl_i$.}
    \label{fig:anneaux}
\end{figure}

\proof{\textsc{Proof.}} By symmetry on the vertices and because $\frac{n!}{(n-k)!}\leq n^k$,
\begin{align*}
    \mathbb P\left[l_{TSP}(k,n)\leq \alpha\right] &=\mathbb E\left[\mb{1}_{l_{TSP}(k,n)\leq \alpha} \right]\\
    &\leq \mathbb E\left[\sum_{1\leq i_1,\cdots,i_k\leq n \text{ distinct}} \mb{1}(|v_{i_2}-v_{i_1}|+\cdots+|v_{i_k}-v_{i_{k-1}}|\leq \alpha) \right]\\
    &\leq n^k \mathbb E\left[ \mb{1}_{|v_2-v_1|+\cdots+|v_k-v_{k-1}|\leq \alpha} \right].
\end{align*}
We next estimate the last term. Given the position of $v_1$, the probability of having $l_1\leq |v_2-v_1|\leq l_1+dl_1$ is at most $\frac{2\pi l_1}{\mathcal A_\cK}dl_1.$ Similarly, conditionally on $v_1,\cdots v_{k-1}$, the probability of having $l_{k-1}\leq |v_k-v_{k-1}|\leq l_{k-1}+dl_{k-1}$ is at most  $\frac{2\pi l_{k-1}}{\mathcal A_\cK}dl_{k-1}$ (see Fig.~\ref{fig:anneaux} for an illustration for $k=4$). Therefore,
\begin{align*}
     \mathbb E\left[ \mb{1}_{|v_2-v_1|+\cdots+|v_k-v_{k-1}|\leq \alpha} \right]
    &\leq \int_{l_1,\cdots,l_{k-1}\geq 0} \mb{1}_{l_1+\cdots + l_{k-1}\leq \alpha} \left(\frac{2\pi l_1}{\mathcal A_\cK}\right)\cdots \left(\frac{2\pi l_{k-1}}{\mathcal A_\cK}\right) dl_1\cdots dl_{k-1}\\
    &= \left(\frac{2\pi\alpha^2}{\mathcal A_\cK}\right)^{k-1} \mathcal P_{k-1},
\end{align*}
where $\mathcal P_{k-1}:=\int_{l_1,\cdots,l_{k-1}\geq 0} \mb{1}_{l_1+\cdots + l_{k-1}\leq 1} \cdot l_1\cdots l_{k-1} \cdot dl_1\cdots dl_{k-1}$. Now for any $k\geq 2$,
\begin{align*}
    \mathcal P_k &= \int_{l_k=0}^1 l_k \left(\int_{l_1,\cdots,l_{k-1}\geq 0} \mb{1}_{l_1+\cdots + l_{k-1}\leq 1-l_k} \cdot l_1\cdots l_{k-1} \cdot dl_1\cdots dl_{k-1}\right)dl_k\\
    &= \int_0^1 l_k\cdot  (1-l_k)^{2(k-1)} \mathcal P_{k-1}\cdot dl_k
    =\mathcal P_{k-1} \cdot \frac{1}{(2k-1)(2k)}.
\end{align*}
Since $\mathcal P_1 = \frac{1}{2}$, by induction $\mathcal P_k = \frac{1}{(2k)!}$. Putting everything together yields the desired result.
\Halmos \endproof

We are now ready to prove a lower bound on the $k$-TSP.

\proof{\textsc{Proof of the lower bound in Theorem \ref{thm:exact rate kTSP}.}} 
Applying Lemma~\ref{lemma:k-tsp estimate}, we obtain
\begin{align*}
    \mathbb P\left[l_{TSP}(k,n)\leq \epsilon\sqrt{\frac{2}{e^2\pi}}\frac{k-1}{n^{\frac{1}{2}\left(1+\frac{1}{k-1}\right)}}\sqrt{\mathcal A_\cK}\right] &\leq n^{k} \left(\frac{4\epsilon^2 (k-1)^2}{e^2 \cdot n^{\left( 1+\frac{1}{k-1}\right)}}\right)^{k-1} \frac{1}{(2k-2)!}\\
    &\leq \left(\frac{4\epsilon^2 (k-1)^2}{e^2}\right)^{k-1} \frac{1}{2\sqrt{\pi (k-1)}}\left(\frac{e}{2(k-1)}\right)^{2(k-1)}\\
    &= \frac{\epsilon^{2k-2}}{2\sqrt{\pi (k-1)}},
\end{align*}
where we used Stirling's approximation $\sqrt{2\pi} n^{n+\frac{1}{2}}e^{-n} \leq n!\leq e n^{n+\frac{1}{2}}e^{-n}.$ Then,
\begin{align*}
    \mathbb E[l_{TSP}(k,n)]  &= \sqrt{\frac{2}{e^2\pi}}\frac{k-1}{n^{\frac{1}{2}\left(1+\frac{1}{k-1}\right)}}\sqrt{\mathcal A_\cK} \int_0^{\infty} \mathbb P\left[l_{TSP}(k,n)\geq \epsilon \sqrt{\frac{2}{e^2\pi}}\frac{k-1}{n^{\frac{1}{2}\left(1+\frac{1}{k-1}\right)}}\sqrt{\mathcal A_\cK}\right] d\epsilon\\
    &\geq \sqrt{\frac{2}{e^2\pi}} \frac{k-1}{n^{\frac{1}{2}\left(1+\frac{1}{k-1}\right)}}\sqrt{\mathcal A_\cK} \int_0^1 \left(1-\frac{\epsilon^{2k-2}}{2\sqrt{\pi (k-1)}}\right) d\epsilon\\
    &\geq \sqrt{\frac{2}{e^2\pi}}\frac{k-1}{n^{\frac{1}{2}\left(1+\frac{1}{k-1}\right)}}\sqrt{\mathcal A_\cK} \left(1-\frac{1}{6\sqrt{\pi}}\right),
\end{align*}
where in the last inequality, we used $\int_0^1 \frac{\epsilon^{2k-2}}{\sqrt{k-1}} \leq \int_0^1 \epsilon^2=\frac{1}{3}$. The result follows.
\Halmos \endproof

This lower bound improves over the simple rate $\Ocal(k/\sqrt n)$ obtained by using the TSP tour only. In particular, when $k$ is small, we can improve the exponent of the denominator---e.g. for $k=1$ we obtain the rate $\Omega(1/n)$ and for $k=2$ we get a rate $\Omega(1/n^{3/4})$. For $k=\Omega(\log n)$, the term $1/(k-1)$ in the exponent of the denominator can be omitted. Thus, the provided lower bound becomes $\Omega(k/\sqrt n)$, matching the simple upper bound with high probability as shown in the following result.

\begin{figure}
    \centering
    \includegraphics[scale=0.65]{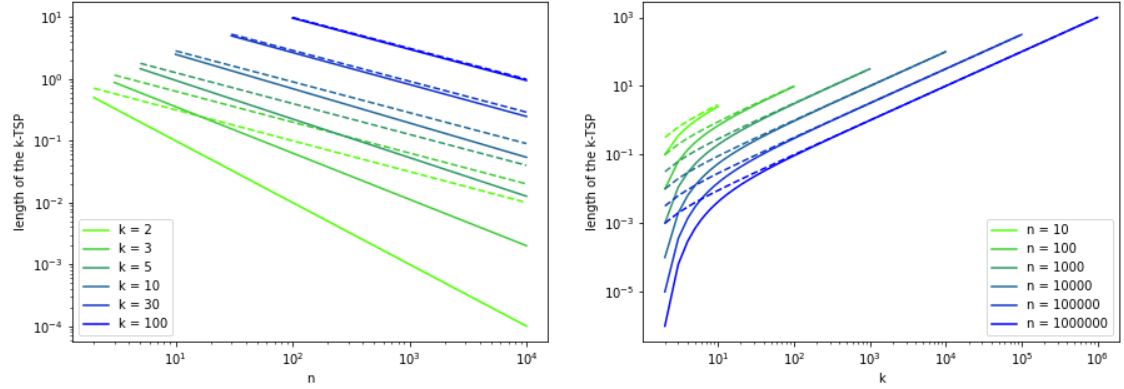}
    \caption{Convergence rate of the length of the $k$-TSP $\Theta\left((k-1)/n^{\frac{1}{2}\left(1+\frac{1}{k-1}\right)}\right)$ (in solid lines) compared to the rate of convergence of the simple heuristic $\Theta((k-1)/\sqrt n )$ (in dashed lines), as a function of $n$ and $k$.}
    \label{fig:my_label}
\end{figure}

\begin{restatable}{corollary}{CorkTSPBound}
\label{cor:k-TSP bound}
Assume all $n$ vertices are drawn independently, uniformly on a compact space $\cK\subset \mathbb R^2$ with area $\mathcal A_\cK$. Denote by $l_{TSP}(k,n)$ the length of the $k$-TSP on these $n$ vertices. Then, there exists a universal constant $M>0$, such that for $M\log n\leq k_n\leq n$,
\begin{equation*}
    \mathbb P\left[l_{TSP}(k_n,n)\leq \frac{k_n}{e\sqrt{\pi n}}\sqrt{\mathcal A_\cK}\right] = o\left(e^{-k_n}\right).
\end{equation*}
\end{restatable}

\proof{\textsc{Proof.}}
We use Lemma~\ref{lemma:k-tsp estimate} with $\alpha = \frac{k_n}{e\sqrt{\pi n}}\sqrt{\mathcal A_\cK}$ and the lower bound $\sqrt{2\pi} n^{n+\frac{1}{2}}e^{-n} \leq n!$ to get
\begin{align*}
    \mathbb P\left[l_{TSP}(k_n,n)\leq \frac{k_n}{e\sqrt{\pi n}}\sqrt{\mathcal A_\cK}\right] \leq n\left(\frac{2k_n^2}{e^2}\right)^{k_n-1} \frac{4k_n^2}{(2k_n)!}\leq  \frac{4e^2n}{\sqrt{\pi k_n}  2^{k_n}}
    \leq \frac{4e^2}{\sqrt{\pi}}\cdot e^{\log n- k_n \log 2}.
\end{align*}
Therefore, for $M>2/\log 2$ and for all $k_n\geq M\log n,$ the right hand side term is $o(e^{-k_n})$.
\Halmos \endproof

\subsection{Upper bound on the \texorpdfstring{$k$}{k}-TSP}
\label{subsection:upper bound kTSP}

In this section, we show that the lower bound shown in Section \ref{subsection:lower bound kTSP} is tight up to a constant factor.

\begin{figure}
    \centering
    \begin{tikzpicture}[scale=0.5]
    \draw[thick]  (0,0) rectangle (9,9) node[right] {$\cK$};
    
    \draw[thick,<->,>=latex] (-0.3,0) -- node[left] {$ 1/m_\alpha$} (-0.3,3);
    \draw[thick,<->,>=latex] (0,-0.3) -- node[below] {$ 1/m_\alpha$} (3,-0.3);
    
    \foreach \y in {3,6}{
        \draw (\y,0) -- (\y,9);
        \draw (0,\y) -- (9,\y);
    }
    \draw (1.683492227580245, 2.389163725360753) node (x1) {$\bullet$};
    \draw (6.351695921256302, 6.692797542574808) node[inner sep=-2] (x2) {$\bullet$};
    \draw (1.1699618836763537, 2.4662987302891097) node (x3) {$\bullet$};
    \draw (8.416704031894087, 7.708190194120887) node[inner sep=-2] (x4) {$\bullet$};
    \draw (7.51021319199524, 3.1499542068114987) node (x5) {$\bullet$};
    \draw (6.721496603453341, 7.790532028026068) node[inner sep=-2] (x6) {$\bullet$};
    \draw (6.114861736477756, 1.9177761758678569) node (x7) {$\bullet$};
    \draw  (7.499930261654734, 1.8711545505630665) node (x8) {$\bullet$};
    \draw  (8.60441285859124, 8.446844457096734) node (x9)[inner sep=-2] {$\bullet$};
    \draw  (5.688328521331934, 5.363653707692948) node (x10) {$\bullet$};
    \draw (0.11560821805908816, 3.8964301349978165)  node (x11) {$\bullet$};
    \draw (1.827244248655064, 1.5905450855339143)  node (x12) {$\bullet$};
    \draw (4.56025554659804, 0.18457866941209269)  node (x13) {$\bullet$};
    \draw  (8.142397925783248, 8.24783199854006)  node[inner sep=-2] (x14) {$\bullet$};
    \draw  (4.5245189678854345, 7.412734753285232)  node (x15) {$\bullet$};
    \draw  (4.463883732550559, 1.422907766587862)  node (x16) {$\bullet$};
    \draw  (7.536118401049598, 3.502715389126722)  node (x17) {$\bullet$};
    \draw (1.6335014699213515, 1.5966283770940446)  node (x18) {$\bullet$};
    \draw (4.266407619051651, 6.32179136547604)  node (x19) {$\bullet$};
    \draw (0.5229135330706505, 6.836150771413925)  node (x20) {$\bullet$};
    \draw (7.668510491401989, 6.320724575385379) node[inner sep=-2]  (x21) {$\bullet$};
    \draw (6.79975226211777, 7.145976467762446) node[inner sep=-2] (x22) {$\bullet$};
    \draw (6.2, 8.6) node[inner sep=-2] (x23) {$\bullet$};
    \draw[thick] (x21) -- (x2) -- (x22) -- (x6) -- (x4) -- (x14) -- (x9) ;
    
    \end{tikzpicture}
    \caption{Illustration of the proof of the upper bound of Theorem \ref{thm:exact rate kTSP} for $n=20$ and $k=7$. The procedure partitions the square into sub-squares then performs the TSP on $k$ points in a sub-square containing at least $k$ points.}
    \label{fig:k-tsp upper bound}
\end{figure}
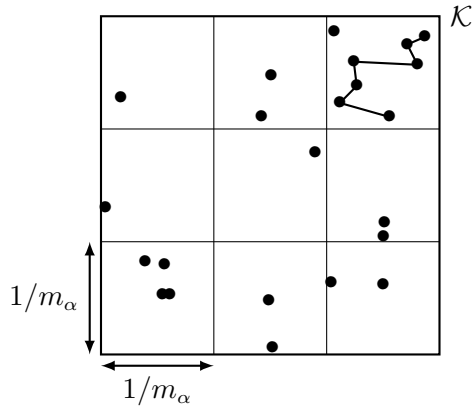

\proof{\textsc{Proof of the upper bound of Theorem \ref{thm:exact rate kTSP}.}}
We first suppose $k\leq n^{1/3}$ and will treat the case $k\geq n^{1/3}$ separately. Fix $\alpha>0$. We start by covering the compact  $\cK$ into $P_\alpha$ disjoint sub-squares of equal size $\frac{1}{m_\alpha}\times \frac{1}{m_\alpha}$ where $m_\alpha:= \left\lfloor \frac{1}{\alpha} \sqrt{n^{1+\frac{1}{k-1}}/(\mathcal A_\cK (k-1))}\right\rfloor$. Because $\cK$ is measurable and has area $\mathcal A_\cK$, we know that $P_\alpha \sim \mathcal A_\cK m_\alpha^2$ as $\alpha\to\infty$. We first show that with high probability, there exists at least one of these sub-squares that contains at least $k$ vertices, and we upper bound $l_{TSP}(k,n)$ by the length of the TSP tour in that sub-square (see Figure~\ref{fig:k-tsp upper bound}). Define $X_i^\alpha$ as the number of vertices in sub-square $i$, for $1\leq i\leq P_\alpha$. Then, $(X_1^\alpha,\cdots,X_{P_\alpha}^\alpha)$ follows a multinomial distribution with $n$ trials and uniform probabilities $1/P_\alpha$. Denote by $A_i^\alpha=\{X_i^\alpha\geq k\}$ the event that sub-square $i$ contains at least $k$ vertices. For any $1\leq i\leq P_\alpha,$ using the fact that $P_\alpha = o(1/n)$,
\begin{align*}
    \mathbb P(A_i^\alpha) = \mathbb P(A_1^\alpha) &\geq \binom{n}{k}\frac{1}{{P_\alpha}^k}\left(1-\frac{1}{P_\alpha}\right)^{n-k}\\
    &\geq \frac{1}{k!} \cdot \frac{n^k}{{P_\alpha}^k} \cdot (1+o(1))\\
    &\geq \frac{(1+o(1))^k(k-1)^k}{k!}\frac{\alpha^{2k}}{n^{1+\frac{1}{k-1}}} \cdot(1+o(1))\\
    &\geq c\cdot  \frac{\alpha^{2k-2}}{P_\alpha},
\end{align*}
for some constant $c>0$. Then, the Bonferroni-Mallows bound for multinomials \citep{mallows1968inequality} implies
\begin{align*}
    \mathbb P\left[\bigcup_{i=1}^{P_\alpha}A_i^\alpha\right] &= 1-\mathbb P[X_1^\alpha\leq k-1,\cdots, X_{P_\alpha}^\alpha\leq k-1]\\
    &\geq 1-\prod_{i=1}^{P_\alpha}\mathbb P(X_i^\alpha\leq k-1)\\
    &\geq 1-e^{-\sum_{i=1}^{P_\alpha} \mathbb P(A_i^\alpha)} \geq 1-e^{-c\cdot \alpha^{2k-2}}.
\end{align*}

Now assume that the event $\bigcup_{i=1}^{P_\alpha} A_i^\alpha$ is met. Let $1\leq i\leq P_\alpha$ the index of a sub-square which contains at least $k$ vertices. Then, according to Lemma~\ref{lemma:upper bound tsp}, the length of the TSP on any $k$ vertices in this sub-square of size $\frac{1}{m_\alpha}\times \frac{1}{m_\alpha}$ is at most $(2\sqrt k+C)/m_\alpha \leq \tilde C \alpha(k-1)\sqrt{\mathcal A_\cK}/n^{\frac{1}{2}\left(1+\frac{1}{k-1}\right)}$ for some universal constant $\tilde C>0$. Therefore, using the previous equation, we get
\begin{equation*}
    \mathbb P\left[l_{TSP}(k,n)> \tilde C\alpha \frac{k-1}{n^{\frac{1}{2}\left(1+\frac{1}{k-1}\right)}}\sqrt{\mathcal A_\cK}\right] \leq 1- \mathbb P\left[\bigcup_{i=1}^{P_\alpha} A_i^\alpha\right] \leq e^{-c\cdot \alpha^{2k-2}}.
\end{equation*}
Finally, we apply the above inequality to obtain
\begin{align*}
    \mathbb E[l_{TSP}(k,n)] &\leq \tilde C\frac{k-1}{n^{\frac{1}{2}\left(1+\frac{1}{k-1}\right)}}\sqrt{\mathcal A_\cK} + \int_{\tilde C(k-1)/n^{\frac{1}{2}\left(1+\frac{1}{k-1}\right)}\sqrt{\mathcal A_\cK}}^\infty \mathbb P[l_{TSP}(k,n)>x]dx\\
    &\leq \tilde C\frac{k-1}{n^{\frac{1}{2}\left(1+\frac{1}{k-1}\right)}}\sqrt{\mathcal A_\cK}\left(1+\int_{1}^\infty e^{-c\cdot \alpha^{2k-2}} d\alpha \right) \leq \hat C\frac{k-1}{n^{\frac{1}{2}\left(1+\frac{1}{k-1}\right)}}\sqrt{\mathcal A_\cK},
\end{align*}
for some universal constant $\hat C$. This ends the proof for $k\leq n^{1/3}$. Now consider the case $k\geq n^{1/3}$. In this case, $n^{\frac{1}{2}\left(1+\frac{1}{k-1}\right)}\sim \sqrt n,$ hence the result can be derived from TSP bounds: let $l_k^*$ be the minimum length of a sub-path of the optimal TSP tour with $k$ consecutive vertices. Since the average length of a path visiting $k$ consecutive vertices is exactly $\frac{k-1}{n}l_{TSP} $, Theorem~\ref{thm:BHH} yields directly $\mathbb E[l_{TSP}(k,n)]\leq \mathbb E[l_k^*] \lesssim\frac{k-1}{n} \beta_{TSP}\sqrt{n\mathcal A_\cK}.$
\Halmos \endproof

The proof of the upper bound is constructive and therefore gives a simple algorithm reaching this bound: first partition the unit square into $P_\alpha$ equal sub-squares, select a sub-square with at least $k$ points, then perform the TSP on any $k$ points in this sub-square (see Fig.~\ref{fig:k-tsp upper bound}). There exists such a sub-square with very high probability. To obtain a constant-factor approximation, we only need a constant-factor approximation of the TSP in the sub-square. For instance, we can use the simple procedure from Lemma~\ref{lemma:upper bound tsp} to obtain a path of length at most $(2\sqrt k + \Ocal(1))/\sqrt{P_\alpha}$. The above procedure may fail to produce a path if no sub-square contains $k$ points, but one can repeat the procedure successively for $\alpha=1,2,3\ldots$ until we find a sub-square with at least $k$ points. By Theorem~\ref{thm:exact rate kTSP}, this algorithm is a constant-factor approximation to the $k$-TSP in expectation.

\subsection{Generalization to non-uniform distributions}
\label{subsection:kTSP non uniform}
Theorem \ref{thm:exact rate kTSP} may be generalized to the case where point positions are drawn independently according to some distribution with a density $f$. For simplicity, we suppose that the density is continuous but the result can be extended to more general densities via smoothing techniques (e.g. Lebesgue derivatives); this is detailed in a companion report \cite{blanchard2022additional}. Because the density is continuous, we can focus on the region of maximum density $\|f\|_\infty$, and relate the $k$-TSP on $n$ points sampled with $f$ to the $k$-TSP on $\|f\|_\infty n$ points sampled uniformly. Hence, we expect the guarantees of Theorem \ref{thm:exact rate kTSP} to hold, replacing $n$ with $\|f\|_\infty n$.

\begin{proposition}
\label{prop:kTSP non uniform}
Assume $n$ vertices are drawn independently, on a compact space $\cK$, according to a continuous density $f$. Denote by $l_{TSP}(k,n)$ the length of the $k$-TSP on these $n$ vertices, where $2\leq k\leq n$. There exists a universal constant $c>0$ such that
\begin{equation*}
    \liminf_{n\to\infty} \mathbb E[l_{TSP}(k,n)] \frac{(\| f\|_\infty n)^{\frac{1}{2}\left(1+\frac{1}{k-1}\right)}}{k-1} \mathcal A_\cK^{\frac{1}{2(k-1)}}\geq c,
\end{equation*}
Further, if $k/n\to 0$ and $k\to\infty$ as $n\to\infty$, there exists a universal constant $C>0$ such that
\begin{equation*}
    \limsup_{n\to\infty} \mathbb E[l_{TSP}(k,n)] \frac{(\| f\|_\infty n)^{\frac{1}{2}\left(1+\frac{1}{k-1}\right)}}{k-1}\leq C.
\end{equation*}
\end{proposition}

\proof{\textsc{Proof.}}
For the lower bound, we use a standard sample-and-reject argument to upper sample the $n$ points according to $f$, from the uniform density on $\cK$ as follows. Consider a sequence $(X_i)$ of i.i.d. uniformly drawn points. A point $X_i=x_i$ is rejected independently of the other points, with probability $1-f(x_i)/\|\tilde f\|_\infty$. The sequence $(Y_i)$ is i.i.d. distributed according to $f$. Using the Hoeffding inequality we show that with probability $1-e^{- n^2/2}$, from $N:=\lceil2\|\tilde f\|_\infty \mathcal A_\cK n\rceil$ uniform draws $(X_i)_{i\leq N}$, at least $n$ points are drawn according to density $f$ with the rejection process. On this event, we lower bound the $k-$TSP length on $n$ points drawn according to $f$ with the $k-$TSP length on the $N$ vertices $(X_i)_{i\leq N}.$ Therefore, using Theorem~\ref{thm:exact rate kTSP}, for some constant $\tilde c>0$,
\begin{align*}
    \mathbb E[l_{TSP(f)}(k,n)] \geq (1-e^{- n^2/2}) \mathbb E[l_{TSP(\mathcal U)}(k,N)]\gtrsim \frac{\tilde c}{2} \frac{k-1}{(\|\tilde f\|_\infty n)^{\frac{1}{2}\left(1+\frac{1}{k-1}\right)}}\mathcal A_\cK^{-\frac{1}{2(k-1)}}.
\end{align*}
Therefore we obtain the desired lower bound. For the upper bound, since $f$ is continuous, there exists a non-empty square $U$ such the density is at least $\|f\|_\infty/2$ on $U$. By the Hoeffding inequality, with probability at least $1-e^{-\epsilon^2 \frac{\|f\|_\infty^2}{2} \mathcal A_U^2 n}$, at least $n_U=\frac{\|f\|_\infty}{2}\mathcal A_U (1-\epsilon)n$ points fell in $U$. Denote by $E_0$ this event on which, these $n_U$ vertices are drawn uniformly on $U$. Then, using Theorem \ref{thm:exact rate kTSP},
\begin{equation*}
    \mathbb E[l_{TSP(f)}(k,n)] \leq n \text{diam}(\mathcal A_\cK)\mathbb P[E_0^c] + \mathbb E[l_{TSP(U)}(k,n_U)] \leq   (1+o_n(1))\cdot C\frac{k-1}{n_U^{\frac{1}{2}\left(1+\frac{1}{k-1}\right)}}\sqrt{\mathcal A_U}.
\end{equation*}
Because $\mathcal A_U^{\frac{1}{k-1}}=O(1)$, the desired upper bound follows.
\Halmos\endproof

The intuition of this generalization is fairly simple: instead of solving the $k-$TSP on the whole compact space $\cK$, we can focus on zones where the density is maximal. The hypothesis $k_n=o(n)$ ensures that this restriction is feasible (otherwise, there would not be $k_n$ points locally). When $k=o(n)$ and $k\to \infty$, the proposed local strategy---performing the $k-$TSP on the highest-density zone---is constant-factor optimal in expectation. As suggested by Proposition~\ref{prop:kTSP non uniform}, this is not exactly the case when $k=O(1)$, for which restricting to a fixed high-density zone affects the local concentration property of the large-deviations analysis.

\section{The Traveling Repairman Problem}
\label{section:TRP}

We now turn to the TRP, which seeks a tour minimizing total latency (Equation~\eqref{eq:total_latency}). For simplicity, assume that we can chose any point as the starting point. Indeed, we will show that the TRP objective is $\Theta(n\sqrt n)$, while an initial edge from a fixed depot to any starting point only affects the TRP objective by an additive $\Ocal(n)$ term.

To provide intuition on the rate $\Theta(n\sqrt n)$, assume that the points are sampled uniformly on a compact space. For the $k$-th served point of the TRP tour with $k\geq k^*=\lfloor n/2\rfloor$, we have $l_k\geq l_{TSP}(k^*,n)$. Then, by Theorem \ref{thm:exact rate kTSP}, the expected latency of the $k$-th point is $\Omega(\sqrt n)$. Because this holds for all $k\geq n/2$, the expected total latency is $\Omega(n\sqrt n)$. Similarly, we can give a simple argument for an upper bound of the expected TRP objective. Consider following the optimal TSP tour of length $l_{TSP}$ with a starting point chosen uniformly at random among the $n$ points. Since the position of each vertex in the tour is uniform, Eq~\eqref{eq:total_latency} implies that the expected latency is equal to $\frac{n-1}{2}l_{TSP}=\Ocal(n\sqrt n)$. Therefore, the expected TRP objective is $\Theta(n\sqrt n)$ for the uniform distribution.

Let us now turn to the case of a general distribution. We show that the TRP objective is still $\Theta(n\sqrt n)$ but we specify the dependence of the constant on the sampling distribution. We state the main asymptotic result which we prove in the following two subsections.

\begin{theorem}
\label{thm:asymptotic bound TRP}
Assume all $n$ vertices are drawn according to a distribution with density $f$ on a compact space $\cK\subset \mathbb R^2$. Denote by $l_{TRP}$ the optimal TRP objective of a tour. Then,
\begin{equation*}
    c\iint_{\cK^2} g_f(x,y) dx dy \leq \liminf_{n\to\infty} \frac{\mathbb E\left[ l_{TRP}\right]}{n\sqrt n} \leq \limsup_{n\to\infty} \frac{\mathbb E\left[ l_{TRP}\right]}{n\sqrt n} \leq C\iint_{\cK^2} g_f(x,y) dx dy
\end{equation*}
where $0<c<C$ are two universal constants and
\begin{equation*}
    g_f(x,y) =  f(y) \left(\mb{1}_{f(y) < f(x)} + \frac{1}{2}\cdot\mb{1}_{f(y) = f(x)}\right) \sqrt{f(x)}.
\end{equation*}
\end{theorem}

\subsection{Lower bound on the TRP}
\label{subsection:lower bound TRP}
We first prove the lower bound of Theorem~\ref{thm:asymptotic bound TRP}. To do so, we approximate the densities as piece-wise constant on sub-squares of the compact space $\cK$. We begin with the case of distributions on the unit square $[0,1]^2$ with piecewise-constant density of the form
\begin{equation}\label{eq:piece_wise_constant_definition}
    f(x)=\sum_{1\leq k\leq m^2} f_k \mb{1}_{Q_k}(x),
\end{equation}
where $\{Q_i\}$ is the regular partition of the unit square into $m^2$ sub-squares of side $1/m$. Note that since $f$ is a density, $\sum_{k=1}^{m^2} f_k = m^2$. We denote by $f_* = \min \{f_k, f_k>0\}$ the minimum positive density across sub-squares. By construction, sampling a vertex from density $f$ is equivalent to choosing one of the squares, with a probability $\frac{f_k}{m^2}$ associated to square $Q_k$, then choosing a point at random uniformly in the chosen $Q_k$. Let $N_k=|\{i, v_i\in Q_k\}|$ denote the number of points in each sub-square. By the strong law of large numbers, we know that $\frac{N_k}{n}\to \frac{f_k}{m^2}$ almost surely.

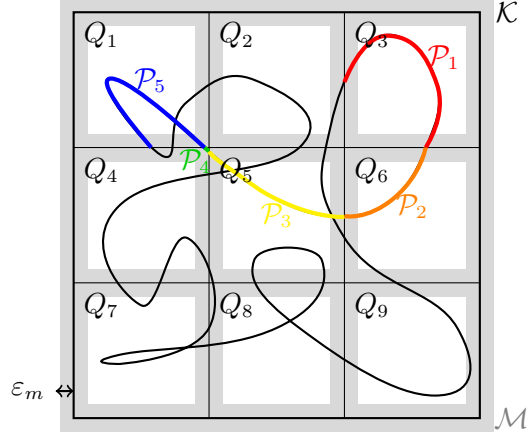
\begin{figure}
    \centering
    \begin{tikzpicture}[scale = 0.6]
    \foreach \y in {0,3,6,9}{
        \fill[gray!30] (\y-0.3,-0.3) rectangle (\y+0.3,9.3);
        \fill[gray!30] (-0.3,\y-0.3) rectangle (9.3,\y+0.3);
    }
    \draw[gray] (9.7,0) node {$\mathcal{M}$};
    
    \draw[thick]  (0,0) rectangle (9,9);
    \draw (9.6,9) node {$\cK$};
    \foreach \y in {3,6}{
        \draw (\y,0) -- (\y,9);
        \draw (0,\y) -- (9,\y);
    }
    \draw[thick,<->] (0,0.6) -- (-0.4,0.6) node[left] {$\epsilon_m$};
    
    \draw[thick] plot [smooth,tension=0.9] coordinates{
    (1,1.2) (3,2) (2.5,4) (1.5,2.5) (1,4.7) (5,6) (4,7.5) (2.5,7) (2,5.8) (1,7.5) (5.5,4.5) (8,6.5) (7,8.5) (5.7,6.5) (6,4) (8.5,2) (8,0.5) (5,2) (4,3.5) (5.5,3.5) (4.5,2) (1,1.4) (1,1.2)
     };
    
    \draw[ultra thick,red] plot [smooth,tension=1] coordinates{(6,7.45) (6.95,8.48) (8.05,7.4) (7.8,6)};
    \draw[ultra thick,red] (8.2,8) node {$\mathcal P_1$};
    
    \draw[ultra thick,orange] plot [smooth,tension=1.2] coordinates{(7.8,6) (7.1,4.88)  (6,4.48) };
    \draw[ultra thick,orange] (7.5,4.7) node {$\mathcal P_2$};
    
    \draw[ultra thick,yellow] plot [smooth,tension=1] coordinates{(6,4.46) (4.8,4.72) (3.5,5.5)  (3,5.92) };
    \draw[ultra thick,yellow] (4.5,4.5) node {$\mathcal P_3$};
    
    \draw[ultra thick,black!20!green] plot [smooth,tension=1] coordinates{(3,5.92) (2.9,6) };
    \draw[ultra thick,black!20!green] (2.7,5.7) node {$\mathcal P_4$};
    
    \draw[ultra thick,blue] plot [smooth,tension=1.4] coordinates{(2.9,6) (0.83,7.53) (1.72,6) };
    \draw[ultra thick,blue] (1.7,7.5) node {$\mathcal P_5$};

    \draw (0,9) node[below right] {$Q_1$};
    \draw (3,9) node[below right] {$Q_2$};
    \draw (6,9) node[below right] {$Q_3$};
    \draw (0,6) node[below right] {$Q_4$};
    \draw (3,6) node[below right] {$Q_5$};
    \draw (6,6) node[below right] {$Q_6$};
    \draw (0,3) node[below right] {$Q_7$};
    \draw (3,3) node[below right] {$Q_8$};
    \draw (6,3) node[below right] {$Q_9$};

    \end{tikzpicture}
    \caption{Illustration of the partition procedure of a TRP tour into sub-paths $\mathcal P_1,\ldots, \mathcal P_P$ corresponding to the partition of the unit square $\cK=[0,1]^2$ into sub-squares $Q_k$ for $1\leq k\leq m^2$ ($m=3$ here). The margin $\mathcal M$ is represented in grey. A sub-path $\mathcal P_i$ in a sub-square $Q_k$ that crosses completely the margin has length at least $\epsilon_m$. We can then lower bound the length of that sub-path in terms of number of visited vertices, using Lemma~\ref{lemma:lower bound length subpath}.}
    \label{fig:sub-paths and margin}
\end{figure}

Now, consider the optimal TRP tour. We would like to restrict the problem on each of the sub-squares. To do so, we can partition the tour into sub-paths such that each sub-path is contained completely in a sub-square $Q_k$ (see Fig.~\ref{fig:sub-paths and margin}). However, unlike for the TSP, we cannot ``glue" the sub-paths in a same sub-square $Q_k$ directly together because here the order of sub-paths impacts the TRP objective. To circumvent this issue, we derive a lower bound of the length of each sub-path individually, in order to obtain a lower bound on the TRP using the results on the $k$-TSP. To minimize the TRP objective, we order sub-paths by decreasing ``vertex density", defined as the ratio between the number of visited vertices in the sub-path and the length of the sub-path.

Define a margin $\mathcal M$ of the borders of the partition $\{Q_k\}$. The margin on each of the sub-squares is set such that any point of $Q_k$ outside of the margin is at a certain distance from the boundary $\partial Q_k$. We will then be able to use Corollary~\ref{cor:k-TSP bound}. More precisely, denote by $B(0,1)$ the unit ball centered at the origin. Define for $\epsilon_m := \frac{\epsilon}{m}$ the margin where $\epsilon>0$ is a arbitrarily small constant:
\begin{equation*}
    \mathcal M = \bigcup_{1\leq k\leq m^2} \left( \partial Q_k + \epsilon_m B(0,1)\right).
\end{equation*}

\begin{restatable}{lemma}{LemmaSmallMargin}
\label{lemma:small margin}
We have $\mathbb P(|V\cap\mathcal M|\geq 8\epsilon n]\leq e^{-\epsilon n}$ where $c>0$ is a constant.
\end{restatable}

\proof{\textsc{Proof.}}
The probability of a vertex falling inside the margin is equal to the area of the margin $\mathcal A_{\mathcal M}$. Then, $\mathcal A_{\mathcal M} \leq 4m\frac{\epsilon}{m} = 4\epsilon$.
Now denote $c=\frac{2}{\sqrt{\pi e}}$. Applying the Chernoff bound to the case of $n$ Bernouilli $\mathcal B(\mathcal A_{\mathcal M})$ samples, we obtain $\mathbb P \left[\left|V\cap \mathcal M \right|\geq 8\epsilon n \right] \leq e^{-\frac{4\epsilon n}{3}}$.
\Halmos \endproof

This lemma shows that the margin only contains a small fraction of vertices. Equivalently, most of the sub-paths in $Q_k$ will visit a vertex in $Q_k\setminus \mathcal M$. These sub-paths have length at least $\epsilon_m$ because they cross the margin completely. Let us now introduce the event $E_0$ as follows,
\begin{equation*}
     E_0= \bigcap_{k\in\{1,\cdots,m^2\}:f_k>0}\left\{\frac{f_k}{2m^2}n\leq N_k\leq \frac{3f_k}{2m^2}n, \quad l_{TSP(Q_k)}\left(\left\lceil \epsilon\cdot e\sqrt{\pi \frac{3f_k}{2m^2}n} \right\rceil,N_k\right)> \epsilon_m\right\},
\end{equation*}
in which we can bound the number of points falling in each sub-squares around their mean $\frac{f_k}{m^2}n$, and lower bound on the maximum number of points that can be visited by a path of length $\epsilon_m$.

\begin{restatable}{lemma}{LemmaProbaE}
\label{lemma:proba E0}
The event $E_0$ has probability $\mathbb P(E_0)=1-o\left(e^{-c\epsilon \sqrt{(f_* n)/m^2}}\right)$ for some constant $c>0$.
\end{restatable}

\proof{\textsc{Proof.}}
By the Chernoff bound, $\mathbb P\left[\left|N_k- \frac{f_k}{m^2}n\right|\geq \frac{f_k}{2m^2}n\right] \leq 2 \exp\left(- \frac{f_k}{12 m^2} n\right).$ Moreover, using Corollary~\ref{cor:k-TSP bound}, we obtain for each $1\leq k \leq m^2$, such that $f_k>0$:
\begin{align*}
    \mathbb P \left[\left. l_{TSP(Q_k)}\left(\left\lceil \epsilon\cdot e\sqrt{\pi \frac{3f_k}{2m^2}n} \right\rceil,N_k\right)\leq \frac{\epsilon}{m} \right|\right.&\left.\frac{f_k}{2m^2}n< N_k< \frac{3f_k}{2m^2}n\right]\\
    &\leq \mathbb P \left[ l_{TSP(Q_k)}\left(\left\lceil\epsilon\cdot e\sqrt{\pi \frac{3f_k}{2m^2}n}\right\rceil,\frac{3f_k}{2m^2}n\right)\leq \frac{\epsilon}{m}\right]\\
    &=o\left(e^{-\epsilon\cdot e\sqrt{(3\pi f_k n)/(2m^2)}}\right).
\end{align*}
Last, we use the union bound to end the proof.
\Halmos \endproof

We now assume that $E_0$ is satisfied, and analyze the length of the TRP. Recall that in sub-square $Q_k$ all paths have length at least $\epsilon_m$, except those included in the margin $\mathcal M$. In particular, we can leverage the upper bound on the number of points of a path of length $\epsilon_m$ provided in the event $E_0$ to give a simple lower bound on the length of any sub-path in $Q_k$ with length at least $\epsilon_m$.

\begin{restatable}{lemma}{LemmaLowerBoundLengthSubpath}
\label{lemma:lower bound length subpath}
Let $p$ be a sub-path in $Q_k$, that has length $l_p\geq \epsilon_m$ and visits $n_p$ vertices. Then, there exists a path of length $\epsilon_m$ in the support of $p$ that visits at least $\frac{\epsilon_m n_p}{2 l_p}$ vertices. Furthermore, on the event $E_0$, for $n$ sufficiently large, $    l_p \geq \frac{n_p}{2e\sqrt{2\pi}\cdot  \sqrt{f_k n}}.$
\end{restatable}

\proof{\textsc{Proof.}}
We subdivide sub-path $p$ in $\lceil l_p/\epsilon_m \rceil$ disjoint portions of length at most $\epsilon_m$. Take the portion that visits most vertices and denote by $n_\epsilon$ that number. In particular, $n_p\leq \lceil l_p/\epsilon_m \rceil n_\epsilon\leq (2l_p/\epsilon_m) n_\epsilon,$ since $l_p\geq \epsilon_m$. Note that in $Q_k$, on the event $E_0$, any path that visits at least $k_0= \epsilon\cdot e\sqrt{(3\pi f_k n)/(2m^2)}$ vertices has length at least $\epsilon_m$. Therefore, $n_\epsilon \leq k_0$. Thus, for $n$ sufficiently large, $n_p\leq 2 l_p (k_0/\epsilon_m) = l_p\cdot 2e\sqrt{2\pi }\cdot  \sqrt{f_k n}.$ The proof follows.
\Halmos \endproof

In particular, Lemma~\ref{lemma:lower bound length subpath} shows that the ``vertex density" of sub-paths in $Q_k$ cannot exceed the ``vertex density" of the TSP on the $N_k$ points in $Q_k$, up to a constant. We can now apply this bound to the length of all sub-paths which are not completely included in the margin $\mathcal M$ in order to lower bound the TRP objective.

\proof{\textsc{Proof of the lower bound of Theorem \ref{thm:asymptotic bound TRP}.}} First consider the case of piece-wise constant densities as defined in Eq~\eqref{eq:piece_wise_constant_definition}. Enumerate the sub-paths $\mathcal P_1,\cdots \mathcal P_P$ which are not included completely in the margin by order of appearance in the TRP path, and denote by $n(\mathcal P_i)$ the number of vertices visited by $\mathcal P_i$. Let $l(\mathcal P_i)$ be the length of $\mathcal P_i$, and $k(i)$ the index of sub-square containing $\mathcal P_i$, i.e. $\mathcal P_i\subset Q_{k(i)}$. Last, let $\tau(v)$ be the latency at point $v\in V$. On the event $E_0$, we can give the following lower bound on the TRP objective by applying Lemma~\ref{lemma:lower bound length subpath} to each of the sub-paths $\mathcal P_i$.
\begin{equation}\label{eq:lower_bound_TRP_on_E0}
    l_{TRP} = \sum_{1\leq i\leq P} \sum_{v\in \mathcal P_i} \tau(v) \geq \sum_{1\leq i\leq P} n(\mathcal P_i) \sum_{1\leq j\leq i-1} l(\mathcal P_j)
    \geq \frac{1}{2e\sqrt{2\pi n}} \sum_{1\leq j\leq P} \frac{n(\mathcal P_j)}{\sqrt {f_{k(j)}}} \sum_{j+1\leq i\leq P} n(\mathcal P_i).
\end{equation}
In order to further lower bound the right term, we use the following lemma which states that the ordering of sub-paths minimizing this objective is exactly the ordering by decreasing density $f_{k(i)}$, which formalizes the intuition that it is advantageous to first serve regions with higher density.

\begin{restatable}{lemma}{LemmaBestOrderingSubpaths}
\label{lemma:best ordering subpaths}
A solution of the following minimization problem $
    \min_{\sigma \in \mathcal S_P} \sum_i \frac{n(\mathcal P_{\sigma(i)})}{\sqrt {f_{k(\sigma(i))}}} \sum_{j>i} n(\mathcal P_{\sigma(j)})$
is given by ordering the sub-paths $\mathcal P_i$ by decreasing order of $f_{k(i)}$.
\end{restatable}

\proof{Proof.}
Denote by $C_\sigma$ the objective of the minimization problem for $\sigma\in\mathcal S_P$. Let $1\leq i<j\leq P$. We will compare $C_\sigma$ and $C_{\tilde \sigma}$ where $\tilde \sigma$ was obtained from $\sigma$ by inserting the $j$-th term in $i$-th position. Formally, $\tilde\sigma(j)=\sigma(i)$, for $i<r\leq j$, $\tilde\sigma(r) = \sigma(r-1)$ and other entries are left unchanged. Then,
\begin{equation*}
    C_{\tilde \sigma}-C_\sigma =n(\mathcal P_{\sigma(j)}) \sum_{i\leq r\leq j-1} n(\mathcal P_{\sigma(r)})\left(\frac{1}{\sqrt {f_{k(\sigma(j))}}} - \frac{1}{\sqrt {f_{k(\sigma(r))}}}\right).
\end{equation*}
Assume that for $i\leq r\leq j-1$ we have $\frac{1}{\sqrt {f_{k(\sigma(j))}}} \leq \frac{1}{\sqrt {f_{k(\sigma(r))}}}$. Then, the objective is decreased when we place $\sigma(j)$ in $i-$th position, $C_{\tilde \sigma} \leq C_\sigma.$ We use this argument to order sequentially the permutation $\sigma$. First take the index $i$ which minimizes $\frac{1}{\sqrt {f_{k(i)}}}$. Let $\sigma^*$ be a permutation such that $\frac{1}{\sqrt {f_{k(\sigma^*(i))}}}$ are in increasing order. We can first place $\sigma^*(1)$ as the first index $\tilde\sigma(1)=\sigma^*(1)$ while decreasing the objective $C_\sigma$. We then place $\sigma^*(2)$ as the second index $\tilde\sigma(2)=\sigma^*(2)$, until we reach the permutation $\sigma^*$ of decrasing order of $f_{k(i)}$. Thus, $C_{\sigma^*}\leq C_\sigma$ and $\sigma^*$ is a minimizer of the problem.
\Halmos \endproof

Let us now give estimates on the right hand of Eq~\eqref{eq:lower_bound_TRP_on_E0}. Denote by $\sigma^*$ the ordering on the sub-squares $Q_k$ such that $f_{\sigma^*(k)}$ is decreasing in $k$. Then, on the event $E_0$,
\begin{align*}
    \sum_{1\leq j\leq P} \frac{n(\mathcal P_j)}{\sqrt {f_{k(j)}}} \sum_{j+1\leq i\leq P} n(\mathcal P_i) &\geq \min_{\sigma \in \mathcal S_P} \sum_{i<j} \frac{n(\mathcal P_{\sigma(i)})}{\sqrt {f_{k(\sigma(i))}}} n(\mathcal P_{\sigma(j)}) \\
    &\geq \sum_{1\leq k<t\leq m^2}\frac{N_{\sigma^*(k)} - |V\cap Q_{\sigma^*(k)}\cap \mathcal M|}{\sqrt{f_{\sigma^*(k)}}} \cdot (N_{\sigma^*(t)} - |V\cap Q_{\sigma^*(t)}\cap \mathcal M|)\\
    &\geq \sum_{1\leq k<t\leq m^2}\frac{N_{\sigma^*(k)}}{\sqrt{f_{\sigma^*(k)}}} N_{\sigma^*(t)} - \frac{2}{\sqrt{f_*}} \sum_{1\leq k,t\leq m^2} N_{\sigma^*(t)}|V\cap Q_{\sigma^*(k)}\cap \mathcal M|\\
    &\geq \frac{n^2}{4m^4} \sum_{1\leq k<t\leq m^2}\sqrt{f_{\sigma^*(k)}} f_{\sigma^*(t)} - \frac{2n|V\cap \mathcal M|}{\sqrt{f_*}},
\end{align*}
where in the last inequality, we used the fact that on $E_0$, $N_k\geq \frac{f_k}{2m^2}n$ for all $1\leq k\leq m^2$, and $f_*=\min\{f_k:f_k>0\}$. By Lemma~\ref{lemma:small margin}, with probability $1-o(\exp(-c\epsilon \sqrt{(f_* n)/m^2}))$, the event $E_0$ is met and $|V\cap\mathcal M|\leq 8\epsilon n$. Denote by $E_1$ this event. Therefore, using Eq~\eqref{eq:lower_bound_TRP_on_E0}, on $E_1$,
\begin{equation}\label{eq:lower_bound_TRP_2}
    l_{TRP} \geq \frac{1}{2e\sqrt{2\pi n}} \left( \frac{n^2}{4m^4} \sum_{1\leq k<t\leq m^2}\sqrt{f_{\sigma^*(k)}} f_{\sigma^*(t)}  -\frac{16\epsilon n^2}{\sqrt{f_*}} \right)
\end{equation}
We will now compare the right term of the above inequality with the integral of $g_f$. Note that
\begin{align*}
    \iint_{\cK^2} g_f(x,y)dx dy
    &= \sum_{1\leq k\leq m^2} \frac{\sqrt{f_{\sigma^*(k)}} }{m^2} \int_{\cK} f(y) \left(\mb{1}_{f(y) < f_{\sigma^*(k)}} + \frac{1}{2}\cdot\mb{1}_{f(y) = f_{\sigma^*(k)}}\right) dy\\
    &= \sum_{1\leq k\leq m^2} \frac{\sqrt{f_{\sigma^*(k)}} }{m^2} \left(\frac{1}{2} \frac{f_{\sigma^*(k)}}{m^2} + \sum_{k<t\leq m^2}\frac{f_{\sigma^*(t)}}{m^2} \right)\\
    &= \frac{1}{2m^2} \int_\cK f(x)^{3/2} dx + \frac{1}{m^4}\sum_{1\leq k<t\leq m^2}\sqrt{f_{\sigma^*(k)}} f_{\sigma^*(t)}.
\end{align*}
The first term in the right-hand side can be made arbitrarily small. Indeed, we can repeat the complete procedure with a finest partition of the unit square $[0,1]^2$ into $(\alpha m)^2$ sub-squares where $\alpha\in \mathbb N^*$. For $\alpha$ sufficiently large, we can get $\frac{1}{\alpha^2 m^2}\int_\cK f(x)^{3/2} dx\leq \delta \iint_{\cK^2} g_f(x,y) dx dy$ for any arbitrarily small $\delta>0$. Then, with this partition we have
\begin{equation*}
    \frac{1}{m^4}\sum_{1\leq k<t\leq m^2}\sqrt{f_{\sigma^*(k)}} f_{\sigma^*(t)} \geq (1-\delta)\iint_{\cK^2} g_f(x,y) dx dy.
\end{equation*}
Therefore, taking $\epsilon<\delta \frac{\sqrt{f_*}}{2^6}\iint_{\cK^2} g_f(x,y) dx dy$, Eq~\eqref{eq:lower_bound_TRP_2} implies that on $E_1$,
\begin{equation*}
    l_{TRP} \geq \frac{1-2\delta}{8 e\sqrt{2\pi}} n\sqrt n \iint_{\cK^2} g_f(x,y) dx dy.
\end{equation*}
We now obtain the desired result,
\begin{equation*}
   \liminf_{n\to\infty}\frac{\mathbb E\left[ l_{TRP}\right]}{n\sqrt n} 
   \geq \liminf_{n\to\infty} \mathbb P[E_1] \cdot \frac{1-2\delta}{8 e\sqrt{2\pi}} \iint_{\cK^2}g_f(x,y) dx dy  \geq \frac{1-2\delta}{8 e\sqrt{2\pi}} \iint_{\cK^2} g_f(x,y) dx dy.
\end{equation*}
This ends the proof for the densities of the form $f(x) = \sum_{k=1}^{m^2}f_k \mb{1}_{Q_k}(x).$ Let us now consider the general case of a distribution on a compact space $\cK$ with both singular part and absolutely continuous part with density $f$. We lower bound the TRP objective by the sum of latencies of points which do not lie in the support of the singular part. With this argument, we can restrict to the case of absolutely continuous distributions with density $f$ without loss of generality. By a scaling argument, we can also suppose without loss of generality that $\cK\subset[0,1]^2$. We need the following lemma to approximate $f$ with a piece-wise constant density, which proof is deferred to Appendix \ref{appendix:technical_lemmas}.

\begin{restatable}{lemma}{LemmaPieceWiseConstantApprox}
\label{lemma:piece-wise constant approx}
Let $f$ be a density on $\cK\subset [0,1]^2$. For any $\epsilon>0$, there exists a density $\phi$ of the form $  \phi(x) = \sum_{1\leq k \leq m^2} \phi_k \mb{1}_{Q_k}(x)$ such that $\|\phi-f\|_1 \leq \epsilon$ and $ \left|\iint_{\cK^2} g_\phi-\iint_{\cK^2}g_f \right| \leq \epsilon.$
\end{restatable}

For any $\epsilon>0$, we use Lemma~\ref{lemma:piece-wise constant approx} to take a density $\phi$ of the same piece-wise constant form as in Eq~\eqref{eq:piece_wise_constant_definition} such that $\|\phi-f\|_1 \leq \epsilon$ and $\left|\iint_{\cK^2}g_\phi- \iint_{\cK^2}g_f \right| \leq \epsilon$. By a coupling argument, we can construct a joint distribution $(X,Y)$ such that $X$ (resp. $Y$) has density $f$ (resp. $\phi$), and $\mathbb P(X\neq Y)\leq 2\int_\cK |\phi(x)-f(x)|dx \leq 2\epsilon.$ Define $n_\epsilon:=|\{i,X_i\neq Y_i\}|$. Then,
\begin{align*}
    l_{TRP(\phi)}:=l_{TRP}(Y_1,\cdots,Y_n) &\leq n[l_{TSP}(Y_i,X_i\neq Y_i) +\sqrt 2] + l_{TRP}(Y_i, X_i=Y_i)\\
    &\leq l_{TRP}(Y_i, X_i=Y_i) + 2n\sqrt{n_\epsilon} + n(C +\sqrt 2),
\end{align*}
where in the second inequality we used Lemma~\ref{lemma:upper bound tsp}. Note that using the Hoeffding inequality, we have $n_\epsilon\leq 3\epsilon n$ with probability at least $1-e^{-2\epsilon^2 n}$. Therefore,
\begin{equation*}
    \frac{\mathbb E\left[ l_{TRP(f)}\right]}{n\sqrt n} 
    \geq \frac{\mathbb E\left[ l_{TRP}(Y_i,X_i=Y_i)\right]}{n\sqrt n}\geq \frac{\mathbb E\left[ l_{TRP(\phi)}\right]}{n\sqrt n}  - 2\sqrt {3\epsilon}+ o(1).
\end{equation*}
We can now use the result proved for density $\phi$.
\begin{equation*}
    \liminf_{n\to\infty}\frac{\mathbb E[l_{TRP}(f)]}{n\sqrt n}
    \geq c \iint_{\cK^2} g_\phi(x,y) dx dy - 2\sqrt{3\epsilon}
    \geq  c \iint_{\cK^2} g_f(x,y) dx dy - c\cdot\epsilon - 2\sqrt{3\epsilon}.
\end{equation*}
This holds for any $\epsilon>0$, hence this ends the proof of the desired TRP objective upper bound.
\Halmos\endproof

\subsection{Upper bound on the TRP}
\label{subsection:upper bound TRP}
The proof of the lower bound of Theorem \ref{thm:asymptotic bound TRP} from Section \ref{subsection:lower bound TRP} suggested a procedure visiting points by zones of decreasing density. We now provide a simple construction of a tour that uses this intuition and shows the upper bound from Theorem \ref{thm:asymptotic bound TRP}.

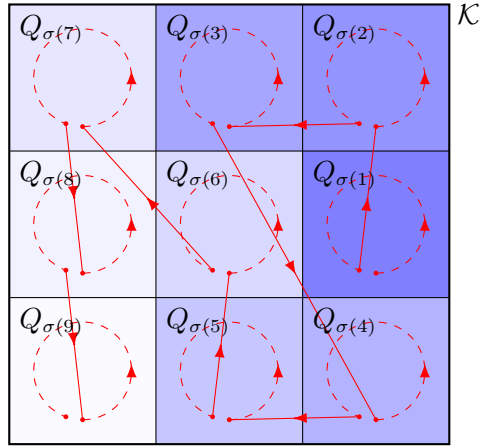
\begin{figure}
    \centering
    \begin{tikzpicture}[scale = 0.65]

    \fill[blue!50] (6,3) rectangle (9,6);
    \fill[blue!42] (6,6) rectangle (9,9);
    \fill[blue!35] (3,6) rectangle (6,9);
    \fill[blue!30] (6,0) rectangle (9,3);
    \fill[blue!22] (3,0) rectangle (6,3);
    \fill[blue!15] (3,3) rectangle (6,6);
    \fill[blue!10] (0,6) rectangle (3,9);
    \fill[blue!5] (0,3) rectangle (3,6);
    \fill[blue!2] (0,0) rectangle (3,3);
    \draw[thick]  (0,0) rectangle (9,9);
    \draw (9.4,8.8) node {$\cK$};
    \foreach \y in {3,6}{
        \draw (\y,0) -- (\y,9);
        \draw (0,\y) -- (9,\y);
    }

    \foreach \y in {0,1,2}{
        \foreach \x in {0,1,2}{
            \draw[dashed, red] (\x*3+1.5,\y*3+0.5) node[inner sep=-5] (B\x\y) {} arc(-90:180+70:1) node[inner sep=-5] (A\x\y) {};
            \draw[ thick,->,>=latex,red] (\x*3+2.5,\y*3+1.5) -- (\x*3+2.5,\y*3+1.65);
            \fill[red] (B\x\y) circle(0.05);
            \fill[red] (A\x\y) circle(0.05);
        }
    }
    
    \draw[red] (A21) -- (B22);
    \draw[red] (A22) -- (B12);
    \draw[red] (A12) -- (B20);
    \draw[red] (A20) -- (B10);
    \draw[red] (A10) -- (B11);
    \draw[red] (A11) -- (B02);
    \draw[red] (A02) -- (B01);
    \draw[red] (A01) -- (B00);
    
    \draw[ thick,->,>=latex,red] ($(A21)!0.5!(B22)$) -- ($(A21)! 0.51!(B22)$);
    \draw[ thick,->,>=latex,red] ($(A22)!0.5!(B12)$) -- ($(A22)! 0.51!(B12)$);
    \draw[ thick,->,>=latex,red] ($(A12)!0.5!(B20)$) -- ($(A12)! 0.51!(B20)$);
    \draw[ thick,->,>=latex,red] ($(A20)!0.5!(B10)$) -- ($(A20)! 0.51!(B10)$);
    \draw[ thick,->,>=latex,red] ($(A10)!0.5!(B11)$) -- ($(A10)! 0.51!(B11)$);
    \draw[ thick,->,>=latex,red] ($(A11)!0.5!(B02)$) -- ($(A11)! 0.51!(B02)$);
    \draw[ thick,->,>=latex,red] ($(A02)!0.5!(B01)$) -- ($(A02)! 0.51!(B01)$);
    \draw[ thick,->,>=latex,red] ($(A01)!0.5!(B00)$) -- ($(A01)! 0.51!(B00)$);

    \draw (0,9) node[below right] {$Q_{\sigma(7)}$};
    \draw (3,9) node[below right] {$Q_{\sigma(3)}$};
    \draw (6,9) node[below right] {$Q_{\sigma(2)}$};
    \draw (0,6) node[below right] {$Q_{\sigma(8)}$};
    \draw (3,6) node[below right] {$Q_{\sigma(6)}$};
    \draw (6,6) node[below right] {$Q_{\sigma(1)}$};
    \draw (0,3) node[below right] {$Q_{\sigma(9)}$};
    \draw (3,3) node[below right] {$Q_{\sigma(5)}$};
    \draw (6,3) node[below right] {$Q_{\sigma(4)}$};

    \end{tikzpicture}
    \caption{Illustration of the constant-factor optimal TRP tour constructed for the upper bound of Theorem \ref{thm:asymptotic bound TRP}. The space is subdivided in sub-squares and the tour performs a constant-factor optimal TSP tour on each of the sub-squares, following the decreasing order of density on the sub-squares. The TSP tour on each sub-square is represented by a dashed path and the density on each sub-square is represented in color---dark (resp. light) blue for high (resp. low) density. Each sub-square is given a priority order from its density: the tour visits zones by decreasing order of density.}
    \label{fig:TRP upper bound}
\end{figure}

\proof{\textsc{Proof of the upper bound of Theorem \ref{thm:asymptotic bound TRP}.}}
By a scaling argument, we suppose without loss of generality that $\cK\subset [0,1]^2$. We use the same notations as in the proof of the lower bound of the expected TRP objective. Let $\epsilon>0$ be a tolerance parameter. Now take $m>0$ and a density $\phi$ given by Lemma~\ref{lemma:piece-wise constant approx} to approximate $f$. We order the sub-squares by decreasing values of $\phi_k$: $\phi_{\sigma(1)}\geq \cdots\geq\phi_{\sigma(m^2)}$. For each of the sub-squares $Q_k$, we construct a tour that is optimal for the TSP --- in practice, only a constant-factor approximation is needed which makes the construction polynomial: one can for example take the tour of Lemma~\ref{lemma:upper bound tsp}. The output TRP tour is given by ``gluing'' together these local TSP tours into a complete tour, following the order $\sigma$. More precisely, we first follow the TSP tour in $Q_{\sigma(1)}$, then the TSP tour in $Q_{\sigma(2)}$ up to the TSP tour in $Q_{\sigma(m^2)}$ (see Fig.~\ref{fig:TRP upper bound}). If a sub-square does not contain vertices we may skip it. As a remark, the additional length for linking the sub-tours is negligible as $n\to\infty$. 

We now prove that this tour is constant-factor optimal with high probability. Define the event $E_0=\bigcap_{1\leq k\leq m^2}\left\{\frac{\phi_k}{2m^2}n\leq N_k\leq \frac{3\phi_k}{2m^2}n\right\},$ where $N_k$ is the count of vertices in sub-square $Q_k$. Recall that $\mathbb E[N_k] = \frac{\phi_k}{m^2}n.$ Therefore, using the same argument as in the proof of lower bound, $E_0$ is met with probability $1-o(\exp(-c\frac{\phi_*}{m^2} n))$, for some constant $c>0$ and where $\phi_*:=\min\{\phi_k:\; \phi_k>0\}$. In the next steps we assume that $E_0$ is met.

By Lemma~\ref{lemma:upper bound tsp}, if we denote by $l_{TSP}^k$ the length of the optimal TSP tour in sub-square $Q_k$, then
\begin{equation}\label{eq:tsp_on_subsquare}
l_{TSP}^k\leq \left(2\sqrt{\frac{3\phi_k}{2m^2}n}+C\right)\frac{1}{m} = \frac{\sqrt{6\phi_k n}}{m^2} + \frac{C}{m}
\end{equation}
for all $1\leq k\leq m^2$ and $C>0$ a universal constant. We are now ready to estimate the TRP objective of our defined tour. Let us denote by $\hat l_{TRP}$ this objective and $\hat l_i$ the distance before visiting vertex $i$ by following the given tour. For each sub-square $Q_k$, denote by $i_k$ the index of the last vertex to be visited in this sub-square by the constructed tour.
\begin{equation*}
\hat l_{TRP} =\sum_{k=1}^{m^2} \sum_{i: v_i\in Q_{\sigma(k)}} \hat l_i
\leq\sum_{k=1}^{m^2} N_{\sigma(k)} \hat l_{i_{\sigma(k)}}
\leq\sum_{k=1}^{m^2}  N_{\sigma(k)} \left( \sum_{l=1}^k l_{TSP}^{\sigma(l)} + (k-1)\sqrt{2}\right).
\end{equation*}
The second term $(k-1)\sqrt{2}$ was obtained by upper-bounding the length of each edge linking a sub-square $Q_{\sigma(l)}$ to the next sub-square $Q_{\sigma(l+1)}$. Therefore, on $E_0$, since $N_k\leq \frac{3}{2}\frac{\phi_k}{m^2}n$ for all $1\leq k\leq m^2$,
\begin{align*}
\hat l_{TRP} &\leq\sum_{k=1}^{m^2}  N_{\sigma(k)} \left( \sum_{l=1}^k l_{TSP}^{\sigma(l)}\right) + \sqrt 2 (m^2-1)\sum_{k=1}^{m^2}  N_{\sigma(k)}\\
&\leq \frac{3n}{2m^2}\sum_{l=1}^{m^2}  l_{TSP}^{\sigma(l)}  \left( \sum_{k=l}^{m^2}\phi_{\sigma(k)}\right) + \sqrt{2}(m^2-1)n\\
&\leq \frac{3}{2}\sqrt{6}\frac{n\sqrt{n}}{m^4}\sum_{1\leq k\leq l\leq m^2} \sqrt{\phi_{\sigma(k)}} \phi_{\sigma(l)}  + \frac{3C}{2m}n + \sqrt{2}(m^2-1)n,
\end{align*}
where in the last inequality we used Eq~\eqref{eq:tsp_on_subsquare}. As in the proof of the lower bound of Theorem \ref{thm:asymptotic bound TRP},
\begin{equation*}
    \frac{1}{m^4}\sum_{1\leq k\leq l\leq m^2}\sqrt{\phi_{\sigma(k)}} \phi_{\sigma(k)}= \iint_{\cK^2}g_\phi(x,y)
    + \frac{1}{2m^2}\int_\cK \phi(x)^{3/2} dx.
\end{equation*}
Therefore, with $\tilde C:=\frac{3\sqrt 6}{2}$, on $E_0$ we obtain
\begin{align*}
    \frac{\hat l_{TRP}}{n\sqrt n} &\leq \tilde C \iint_{\cK^2} g_\phi(x,y) dx dy + \frac{\tilde C}{2m^2}\int_\cK \phi(x)^{3/2} dx + \frac{\sqrt 2 m^2 + 3C/(2m)}{\sqrt n}\\
    &\leq \tilde C\iint_{\cK^2}g_f(x,y) + \tilde C  \left(\epsilon +  \frac{1}{2m^2}\int_\cK \phi(x)^{3/2} dx\right) + \frac{\sqrt 2 m^2 + 3C/(2m)}{\sqrt n}.
\end{align*}
Outside of the event $E_0$, we can use a naive upper bound $\hat l_{TRP}\leq \sum_{i=1}^{n-1}i \sqrt 2\leq\sqrt 2 \frac{n^2}{2}$, obtained by upper bounding the length of each edge by $\sqrt 2$. Since $\mathbb P[E_0^c]=o(\exp(-c\frac{\phi_*}{m^2} n))$, the total contribution of this event is negligible and we obtain
\begin{equation*}
    \limsup_{n\to\infty} \frac{\mathbb E[\hat l_{TRP}]}{n\sqrt n} \leq \tilde C\iint_{\cK^2} g_f(x,y) + \tilde C  \left(\epsilon +  \frac{1}{2m^2}\int_\cK \phi(x)^{3/2} dx\right).
\end{equation*}
Finally, we can take $m$ arbitrarily large, and $\epsilon>0$ arbitrarily small. The result follows.
\Halmos \endproof

We note that this proof of the upper bound uses an ``a priori" algorithm to derive a TRP tour. Namely, the proposed solution visits sub-squares of size $\frac{1}{m}\times \frac{1}{m}$ by decreasing order of density, using only distributional knowledge. Then, the TRP tour is adjusted by visiting the points upon realization of uncertainty, by solving a TSP within each sub-square. This algorithm yields a constant-factor approximation of the optimal TRP latency. As a remark, in order for the estimates in the above proof to hold, we need $m^2\ll n$ for concentration inequalities to hold on the number of points falling in each sub-square. In fact, with similar arguments, one can show that if $\hat l_{TRP}$ denotes the TRP objective obtained by the above procedure, whenever $m^2\ll n$, for any $\epsilon>0$,
\begin{equation*}
     \mathbb P\left[ \hat l_{TRP} \geq (2+\epsilon)n\sqrt n\iint_{\cK^2} g_f(x,y) dx dy \right] \underset{m\to\infty}{\longrightarrow} 0.
\end{equation*}

\section{Fair routing for the \texorpdfstring{$k$}{k}-TSP and the TRP}
\label{section:fairness}

In the first two sections, we provided bounds for the $k$-TSP and the TRP, as well as constant-factor approximation algorithms to provide upper bounds. Both of these approximation schemes rely on a spatial discrimination approach, by prioritizing the zones with high density (high probability density and high point density). Specifically, the approximation scheme for the $k$-TSP visits points only in the highest density zone (Section~\ref{subsection:kTSP non uniform}), and the approximation scheme for the TRP visits zones sequentially by decreasing order of density (Section~\ref{subsection:upper bound TRP}). In fact, these schemes were derived from the lower bound analyses. This suggests that solutions to the $k$-TSP and TRP fundamentally integrate location-based prioritizations.

Therefore, optimizing for the $k$-TSP and the TRP comes at the expense of spatial discrimination. In the proposed $k$-TSP scheme, points that do not lie on the highest density zone will never be visited. Consider the simple setting where a company can choose which customers to serve and generally receives orders from two cities. Following the proposed scheme, the company will exclusively serve customers from the highest-density city and thus ignore customers from one city altogether---even though the densities might be arbitrarily close. Similarly, in the proposed TRP scheme, the waiting time will be much lower in high-density regions than in low-density regions.

To alleviate spatial discrimination outcomes, we incorporate fairness considerations into the $k$-TSP and TRP. Namely, we consider two categories of fairness: (i) geographical fairness, which mitigates disparities across regions, and (ii) population-based fairness, which mitigates disparities across underlying sub-populations. In the aforementioned example, under geographical fairness, the company would need to serve both cities; under population-based fairness, it would need to achieve similar level of service across demographics (based on race or gender, for instance). We quantify the efficiency-fairness trade-off via the \emph{fairness ratio}, defined as the ratio between the objectives of the fair and efficient solutions. This notion relates to the \emph{price of fairness} introduced by \citet{bertsimas2011price}, which measures the relative loss (as compared to the ratio) between the fair and efficient solutions.

\subsection{Fair \texorpdfstring{$k$}{k}-TSP}
\label{subsection:fair kTSP}

We focus on the case $k=o(n)$, $k\to\infty$, and points are sampled according to a continuous density $f$, for which Proposition~\ref{prop:kTSP non uniform} provides an efficient constant-factor algorithm.

\subsubsection{Geographical fairness}
\label{subsubsection:geographical fairness}
Denote by $A_i$ the event where $X_i$ is served. By symmetry, we
focus on the event $A_1$. A first approach to enforce fairness would be to ask that $A_1$ is independent of the position $X_1$. Stated in a more flexible way, we would enforce that the probability of service conditioned on the position exceeds a threshold $\epsilon>0$. We define grographical fairness as follows:
\begin{equation}
\label{eq:geographical fairness}
    \mathbb P(A_1|X_1 = x) \geq \epsilon\frac{k}{n},\quad \forall x\in \cK.
\end{equation}
The discount factor $\frac{k}{n}$ accounts for the fact that only $k$ of the $n$ points can be selected. Indeed, by symmetry, $\mathbb P(A_1) = \frac{1}{n}\mathbb E [\mb 1_{A_1} + \ldots + \mb 1_{A_n}] = \frac{k}{n}$. The minimum service probability imposes to visit the full support of the distribution. This can be viewed as a relaxed version of max-min fairness, in which we would maximize the value of $\epsilon>0$. However, under this requirement, the $k$-TSP loses its locality property, inducing a significant loss in efficiency, formalized in the following proposition.

\begin{proposition}
\label{prop:lower bound geo fair kTSP}
Assume that $\sqrt n\ll k\leq n$. Under geographical fairness (Eq~\eqref{eq:geographical fairness}), the length $l$ of a fair $k$-TSP path satisfies
\begin{equation*}
    \mathbb E[l] \geq (1+o_n(1)) c\cdot\epsilon \frac{k}{\sqrt n} \int_\cK \sqrt f,
\end{equation*}
where $c=\frac{1}{e\sqrt \pi}>0$ is a universal constant.
\end{proposition}

\begin{figure}
    \centering
    \includegraphics[scale=0.55]{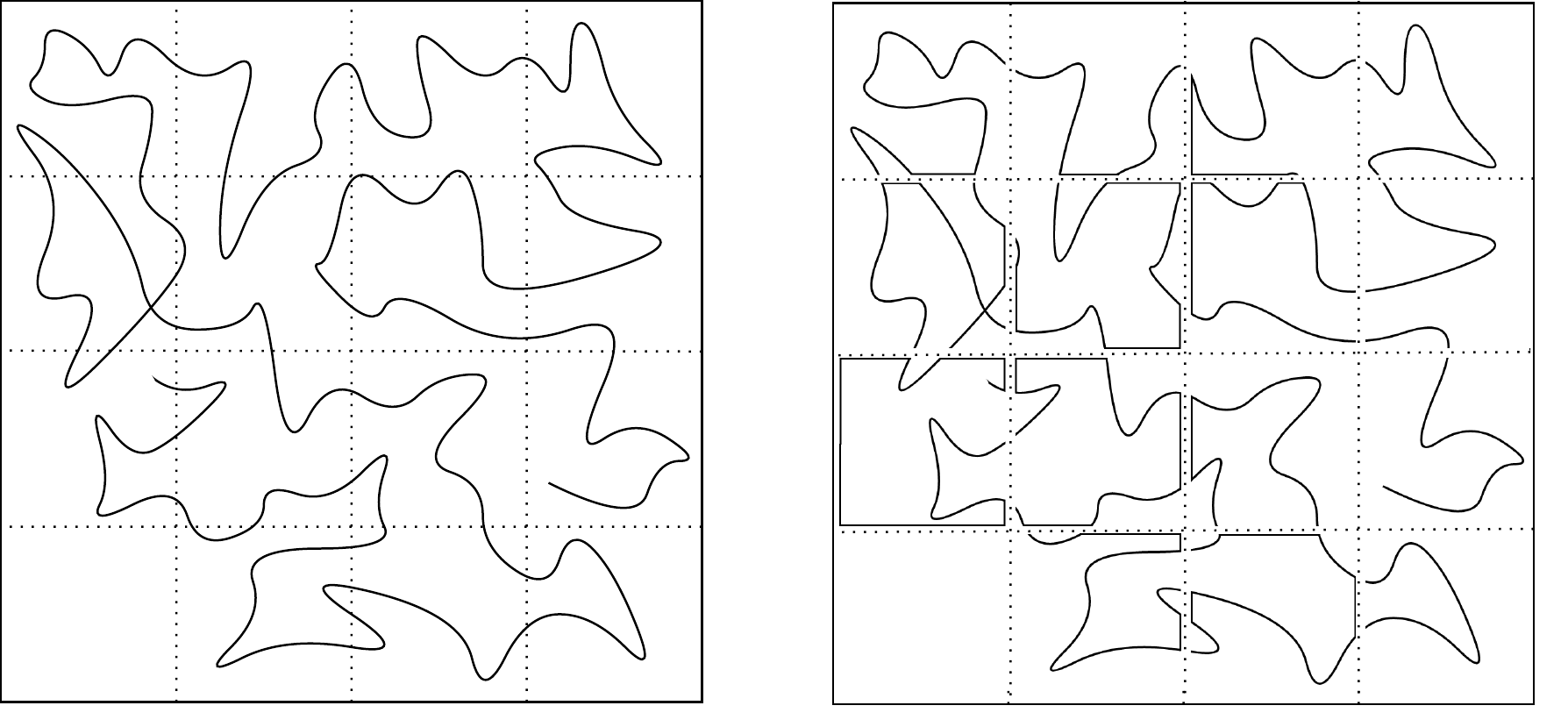}
    \caption{Consider a $k$-TSP path on the left figure. We partition the path into sub-paths in each sub-square, shown in the right figure. The length of the original $k$-TSP path and the sum of length of the sub-paths differs at most by $\Ocal(\mathcal B)$ where $\mathcal B$ denotes the length of the boundary of the partition. In particular, when $k\gg \sqrt n$, the $k$-TSP length grows to infinity. Therefore, the constant boundary length is negligible compared to the $k$-TSP length.}
    \label{fig:classical decomposition on sub-squares}
\end{figure}

\proof{\textsc{Proof.}}
We show the result in the case of distributions on $[0,1]^2$ with piece-wise constant densities on a partition $\{Q_q\}_{q=1}^{m^2}$ defined as in Eq~\eqref{eq:piece_wise_constant_definition}. From a given path visiting $k$ points in the support $[0,1]^2$, we can construct a set of sub-paths in each of the sub-squares such that, together, they visit the same points and have same total length up to a constant dependent only on $m$---the length of the boundary of the sub-squares partition (see Fig.~\ref{fig:classical decomposition on sub-squares}). Since $k\gg\sqrt n$, with high probability, $l_{TSP}(k,n)\geq c\frac{k}{\sqrt n}\to\infty$ for some constant $c>0$. In particular, the additional constant length of the boundary is negligible compared to the length of the path visiting $k$ points. We can now lower bound the length of the path in each sub-square separately. Denote by $n_q$ the number of points visited by the considered path in $Q_q$. and $B_{i,q}$ the event that $X_i$ lies in sub-square $Q_q$. Under fairness constraint (Eq~\eqref{eq:geographical fairness}), we have $\mathbb P[A_i|B_{i,q}]\geq \epsilon\frac{k}{n}$. Then, $\mathbb E[n_q] = n\mathbb E[\mb 1_{A_1\cap B_{1,q}}] \geq n\mathbb P[B_{1,q}]\cdot \epsilon \frac{k}{n} = f_q\cdot \epsilon k.$

Then, if $l_q$ denotes the length of the path reduced to sub-square $Q_q$, we lower bound $l_q$ with the $n_q-$TSP on sub-square $Q_q$ which has at least $\frac{f_k}{m^2}(1-\eta)n$ points with high probability for any fixed $\eta>0$. Using the proof of the $k$-TSP lower bound (Theorem \ref{thm:exact rate kTSP}), with high probability we have
\begin{equation*}
    l_q \geq l_{TSP, Q_q}(n_q,N_q) \geq \frac{n_q\sqrt {1-\eta}}{e\sqrt {\pi f_q n}} - O\left(\frac{\log^2 n}{\sqrt n}\right).
\end{equation*}
Taking the expectation and summing these inequalities yields the desired result on the fair $k$-TSP length, where the $1-o_n(1)$ term corresponds to conditioning on the high probability event.
\Halmos \endproof

As a result, the fairness ratio of a geographically fair $k$-TSP for $k\gg \sqrt n$ compared to the $k$-TSP length (Theorem \ref{thm:exact rate kTSP}), is $ \Omega\left(\sqrt{\| f\|_\infty}\int_\cK \sqrt f\right).$
In Proposition~\ref{prop:lower bound geo fair kTSP} we assumed $k\gg\sqrt n$ for simplicity, but the same non-local behavior would also arise in the general case $k\to\infty$. Essentially, the geographically fair $k$-TSP loses the factor corresponding to the power of choosing which area to serve and the resulting fairness ratio can be arbitrarily large when the density is highly concentrated.

\subsubsection{Population-based fairness}
\label{subsubsection:fairness between populations}
As suggested by the proof of probabilistic bounds in Section~\ref{section:kTSP} and the fairness ratio of geographical fairness, the $k$-TSP is fundamentally spatially unfair. That is, the flexibility to choose which points to visit leads to disregarding zones with low density. Vice versa, imposing to visit all regions with a geographical fairness objective leads to a large loss in efficiency. In response, we now propose a second fairness notion to mitigate the price of fairness.

Consider the setting where points belong to different populations, for instance based on racial demographics, gender demographics, age-based demographics. We aim to design solutions of the $k$-TSP that treat these populations fairly. For instance, one can think of a company constructing an efficient routing procedure while ensuring fairness between distinct sub-populations of customers.

Consider $P$ populations such that points are sampled according to the density $f =f_1+\ldots + f_P$, where $f_i$ corresponds to the distribution of population $i=1,\cdots,P$. For instance, we can view the sampling process as sampling a point according to density $f$, and then assigning population $i$ to this point with probability $\frac{f_i(X)}{f(X)}$. Population-based fairness asks to serve a ``fair'' number of points from each population. We propose deterministic and randomized notions of population-based fairness.

\paragraph{Deterministic population-based fairness}
A natural approach to population-based fairness involves finding a path visiting a fixed proportion $p_i$ of points from each population $i=1,\cdots,P$. For instance, with $p_i = \frac{1}{P}$, this means that the $k$-TSP tour will visit the same number of points from each population; with $p_i = \int f_i$, this means that the $k$-TSP tour will serve each population proportionally to its overall size. However, we will argue that this notion of fairness can be too restrictive and lead to an important loss in terms of efficiency.

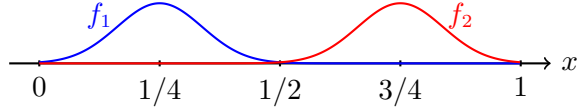
\begin{figure}
    \centering
  \begin{tikzpicture}[scale=0.8]
    \draw[thick, black, ->] (-4.5, 0) -- (4.5, 0) node[right] {$x$};
     
     \draw[thick] (-4, 0.05) -- (-4, -0.05) node[below] {$0$};
     \draw[thick] (4, 0.05) -- (4, -0.05) node[below] {$1$};
     \draw[thick] (-2, 0.05) -- (-2, -0.05) node[below] {$1/4$};
     \draw[thick] (0, 0.05) -- (0, -0.05) node[below] {$1/2$};
     \draw[thick] (2, 0.05) -- (2, -0.05) node[below] {$3/4$};

    \draw[thick, blue] plot[smooth] file {plot1.table};
    \draw[blue] (-3,0.8) node {$f_1$};
    
    \draw[thick, red] plot[smooth] file {plot2.table};
    \draw[red] (3,0.8) node {$f_2$};

  \end{tikzpicture}
    \caption{Case of two populations with truncated Gaussian densities $f_1$ and $f_2$. Under deterministic population-based fairness, the fair $k$-TSP tour needs to visit points where $\min (f_1,f_2)$ is maximal (i.e. at $x=1/2$), instead of visiting points where $f=f_1+f_2$ is maximal ($x\approx 1/4$ or $x\approx 3/4$). In contrast, under randomized population-based fairness, the fair $k$-TSP tour can visit points at $x\approx 1/4$ with probability $1/2$, and points at $x\approx 3/4$ with probability $1/2$. In this example, deterministic population-based fairness yields an arbitrarily large fairness ratio while randomized population-based fairness has an fairness ratio of $1$.}
    \label{fig:gaussians}
\end{figure}

Since the fair $k$-TSP has to visit a fixed proportion of points from each population in the same local area, we can lower bound the length of the fair $k$-TSP by the length of the $(p_i k)-$TSP for density $f_i$ in this local area. In particular, the tour is constrained to visit the zone maximizing the local density of the least-represented population $\min_i f_i$, which leads to the following estimate for the length $l$ of a fair $k$-TSP under deterministic population-based fairness.
\begin{equation*}
    \mathbb E[l]\geq (1+o_n(1))c\cdot \frac{k}{\sqrt{ \|\min_i f_i\|_\infty n}},
\end{equation*}
where $c>0$ is a constant depending only on $P$ and the fixed proportions $p_i$. Further, solving the $k$-TSP locally on the region of maximum minimum-population density $\|\min_i f_i\|_\infty$ achieves this lower bound up to a constant by Theorem \ref{thm:exact rate kTSP}. Hence, the efficiency fairness ratio for deterministic population-based fairness is $  \Theta\left(\sqrt{\frac{\|f\|_\infty}{\|\min_i f_i\|_\infty}}\right).$
When the populations are distributed equally over the space $\cK$, this ratio can be close to one. In contrast, when populations are segregated, this ratio can be arbitrarily large. For instance, consider the simple case of $P=2$ populations with truncated Gaussian densities centered in distant points. In this case, $\|\min f_i\|_\infty$ can be arbitrarily small compared to $\|f\|_\infty$ (see Fig.~\ref{fig:gaussians} for an illustration in one dimension). Further, we can note that if two populations do not have intersecting support, the length of any fair $k$-TSP is $\Omega(1)$, while the length of the $k$-TSP vanishes whenever $k\ll \sqrt n$. Hence, the price of fairness may still be arbitrarily large under deterministic population-based fairness.

\paragraph{Randomized population-based fairness:} In light of these limitations, randomized population-based fairness seeks a \emph{distribution} of $k$-TSP tours, as opposed to a single solution. We only ensure that the $k$-TSP tour visits a fixed proportion $p_i$ of points from each population \emph{in expectation}, but every single $k$-TSP tour may deviate from the proportions $p_i$. Again, $p_i = \frac{1}{P}$ corresponds to equal service (in expectation) and $p_i = \int f_i$ corresponds to proportional service (in expectation). Such randomization allows for more flexibility than deterministic fairness since individual paths of the output distribution can possibly serve populations heterogeneously.

For simplicity, we consider the case where densities $f_i$ are piece-wise constant on a partition $\{Q_j\}$ of the unit square $[0,1]^2$ in $m^2$ sub-squares of equal size $\frac{1}{m}\times\frac{1}{m}$ i.e. $f_i = \sum_{j=1}^{m^2} f_{i,j} \mb 1_{Q_j}$. We can relax this assumption by approximating continuous densities with piece-wise densities on the partition for large enough $m$. However, this simplification will be useful to provide intuition on the proposed randomized population-based fairness scheme. We write the total density as $f=\sum_{j=1}^{m^2} f_j \mb 1_{Q_j}$. Without loss of generality, we can omit sub-squares that do not contain points and assume that the total density is positive $f_j>0$ for all sub-squares $Q_j$. Recall that the condition $k\ll n$ ensures that a path visiting $k$ points can be constructed locally for $n$ large enough. We analyze the randomized approximating scheme in which we select a sub-square $Q_j$ with probability $q_j$ then compute an approximating $k$-TSP path in this sub-square, using the algorithm proposed in Section~\ref{subsection:upper bound kTSP}. By symmetry, the $k$-TSP in sub-square $Q_j$ visits $\frac{f_{i,j}}{f_j}k$ points from population $j$ in expectation. Therefore, the randomized fairness constraint for our scheme imposes that:
\begin{equation}
\label{eq:fairness constraint lp}
    \sum_{j=1}^{m^2}q_j \frac{f_{i,j}}{f_j} = p_i,\quad \forall i=1,\cdots,P.
\end{equation}
By Theorem~\ref{thm:exact rate kTSP}, if $l$ denotes the length of the $k$-TSP path output by the randomized scheme:
\begin{equation}
\label{eq:objective lp}
    \mathbb E[l] = \Theta\left( \sum_{j=1}^{m^2} q_j \frac{k-1}{(f_j n)^{\frac{1}{2}\left(1+\frac{1}{k-1}\right)}}\right),
\end{equation}
where $c>0$ is a universal constant. The optimal set of probabilities $q_j$ can be obtained by solving a simple linear program minimizing the objective (Eq~\eqref{eq:objective lp}) under population-based fairness constraint (Eq~\eqref{eq:fairness constraint lp}) on the probability simplex. We obtain:
\begin{equation*}
     \begin{array}{l@{\quad} l l}
    \min & \displaystyle \sum_{j=1}^{m^2} q_j f_j^{-\frac{1}{2}\left(1+\frac{1}{k-1}\right)}, &\\
    \mathrm{s.t.} &  \displaystyle \sum_{j=1}^{m^2}q_j \frac{f_{i,j}}{f_j} = p_i, &\quad \forall i=1,\cdots,P,\\
    & \displaystyle \sum_{j=1}^{m^2} q_j = 1, &\\
    & q_i\geq 0, &\quad \forall i=1,\cdots,P.
  \end{array}
\end{equation*}
Summing all fairness constraints (Eq~\eqref{eq:fairness constraint lp}) shows that the above linear program contains at most $P$ linearly independent equations. Thus, there exist an optimal probability $q^*$ with at most $P$ positive entries. In other words, instead of visiting all $m^2$ sub-squares, there exists an optimal strategy for the randomized fair scheme visiting at most $P$ different sub-squares.

For instance, consider completely segregated populations, that is, populations with disjoint support. Recall that, in this setting, deterministic population-based fairness has an infinite fairness ratio. This is not the case for randomized population-based fairness. Specifically, under randomized population-based fairness, an optimal strategy consists of choosing one sub-square that maximizes the density $f_i$ for each population, then randomly selecting the sub-square to perform the $k$-TSP, consistently with the fairness constraints (see Fig~\ref{fig:gaussians} for an illustration in one dimension).

We can also add a tolerance $\epsilon\geq 0$ for the fairness by relaxing Eq~\eqref{eq:fairness constraint lp} to
\begin{equation*}
    p_i-\epsilon \leq  \sum_{j=1}^{m^2}q_j \frac{f_{i,j}}{f_j} \leq p_i +\epsilon, \quad \forall i=1,\cdots,P.
\end{equation*}
This constraint yields a new linear program for which there still exists an optimal sparse solution $q^*$ with at most $P+1$ non-zero entries. The $\epsilon$ tolerance acts as a regularization term. When $\epsilon\geq 1$, the corresponding algorithm is blind to the fairness constraints, thus amounting to the $k$-TSP in the case $k\gg 1$: it only visits points in the maximum density sub-square (Section~\ref{subsection:kTSP non uniform}). On the other hand, when $\epsilon=0$, we recover the strict fairness constraint Eq~\eqref{eq:fairness constraint lp}. Denoting by $q^\epsilon$ the optimal probability distribution, the fairness ratio when $1\ll k\ll n$ corresponds to the ratio of the linear program objective for the chosen tolerance parameter $\epsilon$ and the objective for tolerance $1$, i.e. $\Theta\left( \sqrt{\max_j f_j} \sum_{j=1}^{m^2} \frac{
    q^\epsilon_j}{\sqrt{f_j}} \right),$
strictly improving over the fairness ratio for deterministic fairness.

\subsection{Fair TRP}
\label{subsection:fair TRP}

Recall from Section~\ref{subsubsection:geographical fairness} that geographical fairness can result in a significant loss in the objective of the for the $k$-TSP. This stems from the fact that the $k$-TSP is fundamentally local for small $k$ (e.g. $k\ll \sqrt n$). In contrast, the TRP has a global objective and visits all points in the space. In this section we will see that our approximation scheme for the TRP can be adapted to geographical fairness without loss in fairness ratio, in particular under max-min fairness. Additional results for other utility-based notions of fairness are given in the companion report \cite{blanchard2022additional}, in which we show that the approximation scheme for the TRP can be efficiently adapted to account for this notion of fairness. In the game-theoretical setting, max-min fairness yields a Pareto optimal allocation by maximizing the minimum utility that all players derive \citep{bertsimas2011price}. In particular, whenever there exist efficient allocations in which all players have same utility, max-min fairness outputs this equitable allocation. In the case of the TRP, we model the utility of a point by a decreasing function of its latency. In this case, max-min fairness seeks the tour visiting all $n$ points and minimizing the worst latency, i.e. the latency of the point that is visited last. In other words, max-min fairness is equivalent to the TSP, which minimizes the total tour length. We show that our proposed algorithm for the TRP in Section~\ref{subsection:upper bound TRP} is asymptotically optimal for the TSP, hence max-min fair.

\begin{proposition}
The approximation algorithm for the TRP described in Section~\ref{subsection:upper bound TRP} is asymptotically max-min fair. Specifically, let $l(TRP)$ be the maximum point-latency for a TRP tour and $l^*$ be the minimum maximum point-latency i.e. the maximum point-latency of a max-min fair allocation. Then, $\mathbb E[l(TRP)] = (1+o_n(1))\mathbb E[l^*].$
\end{proposition}

\proof{\textsc{Proof.}}
The approximation algorithm for the TRP consists in serving sub-squares sequentially by order of decreasing density. If $\sigma$ denotes this ordering, we first perform the TSP on sub-square $Q_{\sigma(1)}$, then on $Q_{\sigma(2)}$, until $Q_{\sigma(m^2)}$. Note that the total length of the edges linking sub-squares is at most $\Ocal(m^2)=\Ocal(1)$ which is negligible compared to the total length of the tour $\Theta( \sqrt n)$. We can then apply the BHH theorem to each sub-square to obtain
\begin{equation*}
    \mathbb E[l(TRP)] = (1+o_n(1))\beta_{TSP} \sqrt n \int_\cK \sqrt f + \Ocal(1) = (1+o_n(1))\mathbb E[l(TSP)],
\end{equation*}
where $l(TSP)=l^*$ is the length of the optimal TSP tour (hence, a max-min fair tour).
\Halmos \endproof

\section{Conclusion}

In this paper, we gave constant-factor probabilistic estimates for the $k$-TSP and the TRP when points are sampled independently according to a known distribution. Specifically, we showed that the optimal $k$-TSP tour grows at a rate of $\Theta\left(k/n^{\frac{1}{2}\left(1+\frac{1}{k-1}\right)}\right)$, and that the optimal TRP latency grows at a rate of $\Theta(n\sqrt n)$. Moreover, our proofs for the upper bounds are constructive, based on intuitive approximation schemes. For the $k$-TSP, a constant-factor approximation algorithm involves performing a TSP tour in a zone with high point concentration. For the TRP, a constant-factor approximation algorithm involves creating a master ``a priori" tour by visiting zones of decreasing probability density, and then performing a TSP tour within each zone. We also proposed adaptations of these algorithms to capture fairness considerations---namely, randomized population-based fairness for the $k$-TSP and geographical fairness for the TRP. As discussed in Section~\ref{subsec:logistics}, these results can have significant practical implications for the design of transportation and logistics systems where the operator strives to minimize customer wait times or passenger wait times---as opposed to merely minimizing operating costs or travel times.

It is worth noting that we analyzed the $k$-TSP and TRP in the Euclidean plane but the results could be generalized to Euclidean spaces of higher dimension with additional technicality. Furthermore, the upper bound given for the TRP uses the master-tour construction from Lemma~\ref{lemma:upper bound tsp} in order to approximate the TSP locally, which yields a simple ``a priori" algorithm. However, directly using the TSP as subroutine would improve the constant $2$ in the upper bound for the TRP to $\beta_{TSP}$, the constant appearing in the asymptotic length of the TSP. A natural question is whether this constant $\beta_{TSP}$ is tight. This would give an equivalence result of the TRP latency, as opposed to our constant-factor estimates. However, in our analysis, improving the constant of our lower bound for the TRP would require improving the constant of the $k$-TSP lower bound. In particular, this would ask whether for large $k$ (e.g. $k = \Omega( \log n)$), the length of the $k$-TSP is $\sim\beta_{TSP}\frac{k}{\sqrt n}$. We leave this question open for future research. Finally, we refer to the companion report \cite{blanchard2022additional} for additional extensions on the $k$-TSP bounds and the fair TRP.

\section*{Acknowledgments.}
This work was partially supported by the Singapore National Research Foundation through the Singapore-MIT Alliance for Research and Technology (SMART) Centre for Future Urban Mobility (FM). The authors thank Bart van Parys for valuable feedback on the manuscript.

\bibliographystyle{informs2014}
\bibliography{references}

\newpage

\begin{APPENDICES}

\section{Proof of Lemma \ref{lemma:piece-wise constant approx}}
\label{appendix:technical_lemmas}

\LemmaPieceWiseConstantApprox*

\proof{\textsc{Proof.}}
By the Cauchy-Schwartz inequality, $\left\|\sqrt f\right\|_1\leq \sqrt{\|f\|_1}=1$. Let $\epsilon>0$ and $M$ such that $\int_\cK f\mb{1}_{f>M}\leq \epsilon$. Then, we can take a density $\phi_\epsilon$ of the right form such that $\|\phi_\epsilon-f\|_1 \leq \epsilon,\epsilon^{3/2}/M$ and $\left\|\sqrt{\phi_\epsilon}-\sqrt{f}\right\|_1\leq \epsilon.$ We can also choose $\phi_\epsilon$ such that all $\phi_k$ are distinct. For the sake of simplicity, we will write $\phi$ instead of $\phi_\epsilon$ for the next derivations. Again, we have $\left\|\sqrt \phi\right\|_1\leq 1$. First,
\begin{multline*}
    \iint_{\cK^2} g_f (\mb {1}_{|\phi(x)-f(x)|\geq \sqrt \epsilon} +\mb {1}_{|\phi(y)-f(y)|\geq \sqrt \epsilon}) dx dy 
    \leq  \int_\cK  \sqrt{f(x)}\mb {1}_{|\phi(x)-f(x)|\geq \sqrt \epsilon}dx\\ + \int_\cK f(y)\mb {1}_{|\phi(y)-f(y)|\geq \sqrt \epsilon} dy.
\end{multline*}
By Cauchy-Scwartz, $ \int_\cK  \sqrt{f(x)}\mb {1}_{|\phi(x)-f(x)|\geq \sqrt \epsilon}dx \leq \sqrt{\int_\cK \mb {1}_{|\phi(x)-f(x)|>\sqrt \epsilon}dx} \leq \sqrt{\|\phi-f\|_1/\sqrt \epsilon} \leq \epsilon^{1/4}$, where we used Markov's inequality. Also, 
\begin{equation*}
    \int_\cK f(y)\mb {1}_{|\phi(y)-f(y)|\geq \sqrt \epsilon} dy \leq \int_\cK f(y)\mb {1}_{f(y)>M} + M\int_\cK \mb {1}_{|\phi(y)-f(y)|\geq \sqrt \epsilon} dy \leq  \epsilon + M\frac{\|\phi-f\|_1}{\sqrt \epsilon}\leq 2\epsilon.
\end{equation*}
Similarly, we obtain
\begin{align*}
    \iint_{\cK^2} g_\phi (\mb {1}_{|\phi(x)-f(x)|\geq \sqrt \epsilon} +\mb {1}_{|\phi(y)-f(y)|\geq \sqrt \epsilon}) dx dy 
    &\leq  \epsilon^{1/4} + \int_\cK \phi(y)\mb {1}_{|\phi(y)-f(y)|\geq \sqrt \epsilon} dy\\
    &\leq \epsilon^{1/4} + \|\phi-f\|_1 + \int_\cK f(y)\mb {1}_{|\phi(y)-f(y)|\geq \sqrt \epsilon} dy \\
    &\leq \epsilon^{1/4} +3\epsilon.
\end{align*}
It now remains to bound the integral of $g_f-g_\phi$ when $|\phi(x)-f(x)|,|\phi(y)-f(y)|< \sqrt\epsilon$.
\begin{align*}
    & \left|\iint_{\cK^2} (g_f-g_\phi)\mb{1}_{|\phi(x)-f(x)|,|\phi(y)-f(y)|< \sqrt\epsilon} dx dy\right|\\
    &\leq  \iint_{\cK^2}  \left|f(y)\sqrt{f(x)}-\phi(y)\sqrt{\phi(x)}\right|dx dy \\
    &+\left|\iint_{\substack{x,y\in\cK\\|\phi(x)-f(x)|,|\phi(y)-f(y)|< \sqrt\epsilon}} \left(\mb{1}_{\phi(y) < \phi(x)} + \frac{\mb{1}_{\phi(y) = \phi(x)}}{2}-\mb{1}_{f(y) < f(x)} - \frac{\mb{1}_{f(y) = f(x)}}{2}\right) f(y)\sqrt{f(x)}dx dy\right| \\
    &\leq \iint_{\cK^2} f(y)\left|\sqrt{f(x)}-\sqrt{\phi(x)}\right|dxdy + \iint_{\cK^2}|f(y)-\phi(y)|\sqrt{\phi(x)}dx dy\\
    &+\left|\iint_{x,y\in\cK,|f(x)-f(y)|< 2\sqrt \epsilon} \left(\mb{1}_{\phi(y) < \phi(x)} + \frac{\mb{1}_{\phi(y) = \phi(x)}}{2}-\mb{1}_{f(y) < f(x)} - \frac{\mb{1}_{f(y) = f(x)}}{2}\right) f(y)\sqrt{f(x)}dx dy\right| \\
    &\leq 2\epsilon+ \frac{1}{2m^2}\|\phi\|_\infty^{3/2} + \left|\iint_{x,y\in \cK,\; |f(x)-f(y)|< 2\sqrt \epsilon} \left(\mb{1}_{\phi(y) < \phi(x)} -\mb{1}_{f(y) < f(x)} - \frac{\mb{1}_{f(y) = f(x)}}{2}\right) f(y)\sqrt{f(x)}dx dy\right|.
\end{align*}

Now consider the function $ g(z):=\int_\cK \mb{1}_{f(x) = z} dx.$ Note that $0\leq g\leq 1$ and $\sum_{z\geq 0} g(z)\leq 1.$ Therefore, the support of $g$ is countable $Supp(g)=\{z_i;\; i\geq 1\}$. Then, $\iint_{\cK^2} \mb{1}_{f(x) = f(y)} f(y)\sqrt{f(x)}dx dy=\sum_i z_i^{3/2}g(z_i)^2$. We now look at the other terms. First note that
\begin{align*}
    \iint_{f(x)=f(y)=z_i}\mb{1}_{\phi(y) < \phi(x)}dx dy &= \frac{1}{2}\left(\iint_{f(x)=f(y)=z_i}\mb{1}_{\phi(y) < \phi(x)}dx dy+ \iint_{f(x)=f(y)=z_i}\mb{1}_{\phi(y) > \phi(x)}dx dy\right)\\
    &= \frac{g(z_i)^2}{2} - \frac{1}{2}\iint_{f(x)=f(y)=z_i}\mb{1}_{\phi(y) = \phi(x)}dx dy.
\end{align*}
Therefore,
\begin{align*}
    \left|\iint_{f(x)=f(y)} \mb{1}_{\phi(y) < \phi(x)} f(y)\sqrt{f(x)}dx dy \right.&\left.-\frac{1}{2}\sum_i z_i^{3/2}g(z_i)^2\right|
    \\
    &\leq\sum_i \left|\iint_{f(x)=f(y)=z_i} \mb{1}_{\phi(y) < \phi(x)} f(y)\sqrt{f(x)}dx dy - \frac{1}{2}z_i^{3/2}g(z_i)^2\right|\\
    &= \sum_i z_i^{3/2}\left|\iint_{f(x)=f(y)=z_i}\mb{1}_{\phi(y) < \phi(x)}dx dy - \frac{g(z_i)^2}{2}\right|\\
    &\leq  \frac{1}{2}\iint_{\phi(y) = \phi(x)}f(x)\sqrt{f(y)}dx dy.
\end{align*}
Because $\phi_k$ is distinct on each sub-square $Q_k$, by the dominated convergence theorem, the right term vanishes when $m$ grows. Indeed, $\mb 1_{\phi(y) = \phi(x)}\to \mb 1_{x=y}$ as $m\to\infty$, and $\{x=y\}$ is a negligible set. From now, we take $m$ sufficiently large such that the right term is upper bounded by $\delta$. Finally,
\begin{equation*}
    \left|\iint_{\cK^2} g_f-g_\phi \right| \leq 2\epsilon^{1/4} + 7\epsilon   + \frac{1}{2m^2}\|\phi\|_\infty^{3/2}+ \delta+\iint_{\cK^2}\mb{1}_{|f(x)-f(y)|< 2\sqrt \epsilon}\mb{1}_{f(x)\neq f(y)}f(x)\sqrt{f(y)}dx dy.
\end{equation*}
By the dominated convergence theorem, the right term vanishes as $\epsilon\to 0$. Then, taking $0\leq\epsilon\leq \delta$ sufficiently small, then $m$ sufficiently large, we can achieve $\left|\iint_{\cK^2} g_f-g_\phi \right| \leq 2\delta.$ Note that we also have $\|\phi_\epsilon-f\|_1\leq \epsilon\leq \delta$. This ends the proof of the lemma.
\Halmos \endproof

 \end{APPENDICES}

\end{document}